\documentclass[amsmath, superscriptaddress, amssymb, aps, prd, longbibliography,nofootinbib,onecolumn, notitlepage,10pt]{revtex4-2}
\usepackage{graphicx}
\usepackage{enumerate}
\usepackage{blindtext}
\usepackage{natbib}
\usepackage{amsmath}
\usepackage{amssymb}
\usepackage{physics}
\usepackage{wrapfig}
\usepackage{braket}
\RequirePackage[colorlinks,citecolor=blue,urlcolor=magenta,linkcolor=blue]{hyperref}
\usepackage[english]{babel}
\usepackage{esint}
\usepackage{dsfont}
\usepackage{soul}
\usepackage[T1]{fontenc}
\usepackage{cleveref}
\usepackage{lipsum}
\begin{document}

\title{Non-local correlations of a test quantum field in gravitational collapse}
\author{Harkirat Singh Sahota}
\email{harkirat221@gmail.com}
\author{Suprit Singh}
\email{suprit@iitd.ac.in}

\author{Ashish Pandita}
\affiliation{Department of Physics,  Indian Institute of Technology Delhi, Hauz Khas, New Delhi, 110016, India.}

\date{February 2025}


\begin{abstract}
     Quantum correlations across the horizon could be pivotal in unveiling the puzzles surrounding quantum aspects of black holes and Hawking radiation. The peaks in the equal time correlation function are typically attributed to the entangled particle excitations. In this work, we have investigated the evolution of the correlations of a test quantum field on a dynamical background spacetime undergoing gravitational collapse.  In the case of super-critical collapse, as the black hole and its horizon forms, correlated peaks are seen to appear across the horizon, representing an entangled Hawking pair. The outside peak moves away from the horizon as the system evolves, possibly representing outgoing Hawking flux. The implications of these non-local correlations are discussed in light of information paradox, quantum atmosphere and analogue black holes.
\end{abstract}

\maketitle

\section{Introduction}

The semiclassical analysis of Hawking showed that for asymptotic observers black holes are radiating objects, emitting a thermal black body radiation at a temperature inversely proportional to their mass \cite{Hawking1975,Wald:1975kc,Unruh:1976db}. This result brought out by investigating quantum fields in curved spacetimes is considered as a milestone of modern theoretical physics; one that opened up several questions about our understanding of nature at the fundamental level \cite{Birrell:1982ix,Kay:1988mu,FULLING1987135,Wald:1999xu,Visser:2001kq,Helfer:2003va,Padmanabhan:2003gd,Fabbri:2005mw,Padmanabhan:2009vy,Visser:2014ypa,Mathur:2009hf,Raju:2020smc,Harlow_2016,Unruh:2017uaw}. Certain open issues regarding Hawking's result can be succinctly classified as: the observability of Hawking radiation \cite{Unruh:1980cg} and epistemological aspects of the Hawking process \cite{Hawking_1976}. Owing to the weak intensity of Hawking radiation \cite{Unruh:1980cg}, a direct observation of this phenomenon is strikingly difficult. On the second front lies the information paradox, which has captivated researchers for decades \cite{Mathur:2009hf,Harlow_2016,Unruh:2017uaw,Raju:2020smc}, leading to a vast landscape of analyses that tries to address this conceptually profound issue \cite{Page:1993df,Page:1993wv,Kraus:1994by,Chakraborty_2017,Almheiri_2020,Almheiri:2020cfm}. Another related question of interest in this regard is ``{\it where does the Hawking radiation originate?}'' \cite{Unruh_atmosphere,Schutzhold:2008tx,Schutzhold:2010ig,Balbinot:2021bnp,Fontana:2023zqz,Kaczmarek:2023kpn,Giddings:2015uzr,Dey:2017yez,Dey:2019ugf}. This work is aimed at addressing these questions by studying the vacuum correlations of a quantum field in a gravitational collapse scenario.

A promising approach in observing the Hawking phenomenon is through analogue systems \cite{Barcelo:2005fc} that mimic characteristic features of a quantum field propagating on a curved spacetime with a horizon. The oldest and most well-known analogy is of an acoustic black hole (dumb hole) \cite{Unruh:1980cg} where an acoustic horizon appears due to the supersonic flow of the Eulerian fluid and sound waves are dragged away by the flow in the supersonic region and thus cannot propagate back to the subsonic region. This opens up avenues for investigating the quantum aspects of gravitational systems in laboratories \cite{Novello2002-ix}. A number of systems have been proposed as candidates for the experimental verification of the Hawking effect \cite{Visser:1997ux,Jacobson:1998ms,Leonhardt:2000fd,Unruh:2003ss,Giovanazzi:2004pm,Schutzhold:2004tv,Philbin:2007ji,Rousseaux:2007is,Weinfurtner:2010nu,Belgiorno:2010wn,Horstmann_2010,Lombardo_2012,Busch_2014,Jacquet:2022vak,PhysRevLett.130.111501,Berti:2024cut} with the sonic black holes produced from the Bose-Einstein condensate being the most studied example \cite{Garay:1999sk,Barcelo:2001ca,Giovanazzi:2004zv,Lahav:2009wx,Macher:2009nz,Recati:2009ya,Giovanazzi_2011,PhysRevA.85.013621,Finazzi:2013sqa,Boiron_2015,Dudley:2018qpz,Dudley:2020toe,Palan_2022,Anderson:2024fct}. 

The physical picture of the Hawking process involves pair creation where one member is created inside the horizon, while the partner is created outside the horizon, which propagates to the asymptotic observer as thermal radiation \cite{Hotta:2015yla}. Since the particles are entangled, one can expect non-local correlations across the horizon \cite{Schutzhold:2010ig}. For the gravitational case, such correlations do not appear \cite{Balbinot:2021bnp,Balbinot:2023grl} as the Hawking quanta inside the horizon are swallowed by the singularity, whereas the investigation of correlations in a toy model with event horizon does exhibit non-local correlations \cite{Schutzhold:2010ig}. This turns out to be the case for the Bose-Einstein-condensate analogue black holes \cite{Garay:1999sk}, where the particle-partner correlations of the Hawking radiation are manifested by the presence of non-local density-density correlations across the horizon \cite{Balbinot,Carusotto:2008ep,Michel_2016}. In the stationary regime, when the sonic horizon is formed, a stationary peak with negative correlations appears with a support perpendicular to the local correlations along the diagonal $(x=x')$. The spontaneous Hawking radiation and the correlations predicted by numerical simulations in analogue systems have been observed experimentally in acoustic black holes \cite{Steinhauer:2014dra,Steinhauer:2015ava,Steinhauer:2015saa,MunozdeNova:2018fxv,Kolobov:2019qfs,Fabbri_2021}, lending credence to the prediction of the Hawking effect.  Investigation of the behavior of vacuum correlations of a test field in a collapsing geometry (actual gravitational system) vis-\`a-vis the density-density correlations for acoustic black holes will shed further light on the characteristics of Hawking radiation.

As such the Hawking process also leads to the information paradox, where an initial pure state evolves to a mixed state during evaporation and, therefore, is in contradiction with unitary evolution \cite{Hawking_1976}. The correlations of Hawking radiation are expected to hold the key to at least addressing some aspects of this paradox \cite{Braunstein:2009my,Mathur_2011,Alonso-Serrano:2015bcr,Alonso_Serrano_2016,Lochan:PRL2015,Lochan:PRD2016,Modak_2015,Saini_2015}. Non-local correlations of Hawking radiation are expected \cite{Schutzhold:2010ig,Balbinot:2021bnp,Fontana:2023zqz,Kaczmarek:2023kpn} to back up the proposal of a quantum atmosphere of the black hole \cite{Giddings:2015uzr,Dey:2017yez,Dey:2019ugf}, which states that the Hawking particles emerge from a region outside the event horizon located at a distance  $\sim\mathcal{O}(\kappa^{-1})$, where $\kappa$ is the surface gravity at the event horizon. Therefore, correlations of the quantum field in collapsing geometry are multifaceted observables that can reveal various aspects of Hawking radiation and, thus, are the objects of our interest. 


Gravitational collapse of a scalar field is the simplest arena in which the nonlinearities of general relativity can be investigated and understood rigorously \cite{Christodoulou:1986du,Christodoulou:1986zr,Christodoulou:1987vu,Christodoulou:1987vv,Goldwirth:1987nu}. Numerical investigations of this system have uncovered critical behaviour and self-similar aspects of spacetime \cite{Choptuik} and the advances in the numerical relativity have led to a better understanding of the gravitational collapse \cite{Lehner:2001wq,Gundlach:2002sx,Gundlach:2007gc,alcubierre}. Recently, Berczi et. al. \cite{BercziGH,Berczi_JHEP,Berczi_PRD} have developed a code and formalism for studying the gravitational collapse of a quantum scalar field. The collapse of a quantum field shows the critical behavior similar to the classical collapse \cite{Berczi_PRD,Berczi_JHEP} (a similar investigation of the semiclassical collapse with only radial modes can be found in \cite{Guenther}), and the correlations of the background scalar field in the coherent state showed interesting features \cite{Berczi:2024yhb}. As the quantum field itself is sourcing the collapse, the correlations get clumped inside the horizon towards the origin while suppressed outside the horizon. We expect different features for the correlations in the case of a test quantum field. We, thus, adapted their code to include two fields (one test quantum field and the other classical field sourcing the geometry), so as to study the vacuum correlations of a test quantum field propagating on a spherically symmetric background spacetime generated by a self-gravitating massless scalar field, in a fully dynamical scenario.


The paper is organized as follows. We start with the ADM formalism of gravitational collapse sourced by a scalar field in spherically symmetric geometry, discuss the test quantum field in critical collapse and briefly sketch the numerical considerations in \cref{Sec2}. In \cref{GC_BD}, we present the numerical evolution of the background geometry for two cases, where the black hole forms and does not form. Identifying the regimes of interest from the background dynamics, we present the dynamics of the vacuum correlations of the test field for these cases in \cref{QC_TF}. We close with a summary and outlook in \ref{Conclusion}.

\section{Scalar field in a spherically symmetric spacetime}\label{Sec2}
The set up under consideration is of a gravitational collapse sourced by a scalar field along with a test quantum field propagating on this dynamical background geometry. We start with the $3+1$ formulation of general relativity in the spherical symmetry.

\subsection{Gravitational collapse sourced by classical scalar field}
We are interested in the self-gravitational dynamics of a scalar field in a spherically symmetric geometry given by the line element
\begin{align}
    ds^2=-\alpha^2(t,r)dt^2+A(t,r)dr^2+r^2B(t,r)d\Omega^2
\end{align}
where $\alpha(t,r)$ is the lapse function, $A(t, r)$ and $B(t,r)$ determine the geometry of a constant time hypersurface. The goal is to solve the Einstein-Klein-Gordon (EKG) system for this spacetime:
\begin{align}
    G_{ab}=\frac{1}{M_P^2}T_{ab},\quad \Box\Phi_c=0,
\end{align}
where $T_{ab}$ is the stress-energy tensor of the scalar field and $M_P$ is the reduced Planck mass. These are coupled second-order differential equations which can be cast into a system of first-order equations by defining additional variables, viz.,
\begin{align}
    K_A:=-\frac{1}{2\alpha}\frac{\dot{A}}{A},\quad K_B := -\frac{1}{2\alpha}\frac{\dot{B}}{B},\quad\lambda:=\frac{1}{r}\left(1-\frac{A}{B}\right),\quad 
    D_A:=\frac{A'}{A},\quad D_B:=\frac{B'}{B},\quad D_\alpha:=\frac{\alpha'}{\alpha}.
\end{align}
Here, the dot represents the derivatives with respect to $t$ and the prime represents the derivatives with respect to $r$. To ensure strong hyperbolicity of the system for all gauge choices \cite{alcubierre}, it is appropriate to work with $(K,\;\Tilde{U})$ instead of $(K_A,D_A)$ defined as $\Tilde{U}:=D_A-2D_B-4B\lambda/A$ and $K:=K_A+2K_B$ being the trace of the extrinsic curvature.
The non-vanishing components of the Einstein tensor for this geometry then take the form,
\begin{align}
    G^t_{\:\:\: t}=&\frac{1}{A}\Bigg[ D_B'+\frac{1}{r}\Big( \lambda +D_B-\Tilde{U}-\frac{4\lambda B}{A} \Big)
    -D_B\Big( \frac{1}{4}D_B+\frac{1}{2}\Tilde{U}+\frac{2\lambda B}{A}\Big) \Bigg]-K_B(2K-3K_B), \label{Gtt}\\
    G^t_{\:\:\: r}=&\frac{2}{\alpha}\Bigg[ -K_B'+\Big( \frac{1}{r} +\frac{D_B}{2} \Big)(K-3K_B) \Bigg], \label{Grt} \\
    G^r_{\:\:\: r}=&\frac{2}{\alpha}\Bigg[\dot{K}_B-\frac{3}{2}\alpha K_B^2 +\frac{\alpha}{2r^2A}\Big( 1-\frac{A}{B} \Big) +\frac{\alpha}{2rA}D_B +\frac{\alpha}{8A}D_B^2+\frac{\alpha}{rA}D_{\alpha}+\frac{\alpha}{2A}D_B D_{\alpha} \Bigg],\\
    G^{\theta}_{\:\:\: \theta}=&\frac{1}{\alpha}(\dot{K}-\dot{K}_B) 
    -K^2+3K_B(K-K_B) +\frac{1}{2A}(D_B'+2D_{\alpha}')-\frac{1}{4A}\Bigg[(D_B+2D_{\alpha})\Big( 
    \Tilde{U}+D_B+\frac{4B\lambda}{A}\Big)-4D_{\alpha}^2 \Bigg] \notag \\
    &+ \frac{1}{2rA}\Big( 2D_{\alpha}-\Tilde{U}-\frac{4B\lambda}{A} \Big).  
\end{align}
The $tt$ and $tr$ components of the Einstein tensor contain no time derivative, and therefore, the corresponding equations are non-dynamical, that is, these are simply constraint equations called the Hamiltonian constraint and momentum constraint. The dynamical system is a constrained initial value problem, where the physical initial conditions are those that satisfy the constraint equations. A standard approach for solving this system of non-linear equations is to provide the initial data that satisfies the constraints on a constant time slice and evolve the system using dynamical equations. We will return to this later during gauge fixing. The matter sector is comprised of the massless scalar field whose stress-energy tensor is given by
\begin{align}
    T_{ab}=\partial_a\Phi_c\partial_a\Phi_c-\frac{1}{2}g_{ab}\left[g^{ij}\partial_i\Phi_c\partial_j\Phi_c\right].
\end{align}
To cast the dynamical equations for the scalar field in the first-order form, we define the variables,
\begin{align}
    \Pi:=\frac{\sqrt{A}B}{\alpha}\dot{\Phi}_c,\quad\Psi:=\Phi_c'.\label{dphi}
\end{align}

With this choice, the components of the stress-energy tensor are given by
\begin{align}
    \rho&=n^an^bT_{ab}=\frac{1}{2A}\left(\frac{\Pi^2}{B^2}+\Psi^2\right),\\
    j_A&=-n^aT_{ar}=-\frac{\Pi\Psi}{\sqrt{A}B},\\
    S_A&=\gamma^{rr}T_{rr}=\frac{1}{2A}\left(\frac{\Pi^2}{B^2}+\Psi^2\right)=\rho,\\
    S_B&=\gamma^{\theta\theta}T_{\theta\theta}=\frac{1}{2A}\left(\frac{\Pi^2}{B^2}-\Psi^2\right),
\end{align}
where $n^a$ is the normal vector to the equal time hypersurface and $\gamma^{ij}$ is the inverse of the induced metric on the hypersurface. For the gravitational sector, the Einstein's equations
\begin{align}
    G^t_{\:\:\: r}=&-\frac{\alpha j_A}{AM_P^2},\quad G^t_{\:\:\: t}=-\frac{\rho}{M_P^2},\quad G^r_{\:\:\: r}=\frac{S_A}{M_P^ 2},\quad G^\theta_{\:\:\: \theta}=\frac{S_B}{M_P^2}\label{EEs_1st}
\end{align}
can be cast as the first-order equations for the metric variables as 
\begin{align}
    \dot{A} = -2\alpha A (K- 2K_B),\quad\dot{B} = -2\alpha B K_B,\label{dt_AB}
\end{align}
with the additional variables satisfying the equations
\begin{align}
        &\dot{D}_B = -2 \partial_r (\alpha K_B),\quad\dot{\lambda}= \frac{2\alpha A}{B}\frac{(K-3K_B)}{r},\label{dt_DBl}\\
        &\dot{\Tilde{U}}=-2\alpha \Big[K'+D_{\alpha}(K-4K_B)-2(K-3K_B)\Big( D_B-\frac{2\lambda B}{A} \Big)\Big] - 4\alpha \frac{j_A}{M_P^2},\label{dt_Utilda}\\
        &\dot{K}=\alpha(K^2-4KK_B+6K_B^2) -\frac{\alpha}{A}\Big[
        D_{\alpha}'+D_{\alpha}^2+\frac{2D_{\alpha}}{r}-\frac{1}{2}D_{\alpha}
        \Big( \Tilde{U} + \frac{4\lambda B}{A} \Big) \Big]  +\frac{\alpha}{2M_P^2}(\rho + S_A + 2S_B-2\Lambda),\label{dt_K}\\
        &\dot{K}_B=\frac{\alpha}{Ar}\Big[ \frac{\Tilde{U}}{2}+\frac{2\lambda B}{A} -D_B-\lambda-D_{\alpha} \Big] + \frac{\alpha}{A} \Big[
        -\frac{D_{\alpha}D_B}{2}-\frac{D_B'}{2}+ \frac{D_B}{4}\Big( \Tilde{U}+\frac{4\lambda B}{A} \Big)+AKK_B \Big] + \frac{\alpha}{2M_P^2}(S_A-\rho).\label{dt_KB}
\end{align}
The equation of motion for $\lambda$ is well defined in the continuum limit but the numerical evolution can be highly unstable due to the presence of $1/r$ factor. This issue is resolved by using the momentum constraint equation leading to,
\begin{equation}
    \dot{\lambda}=\frac{2\alpha A}{B}\Big[ K_B'-\frac{1}{2}D_B(K-3K_B)+
    \frac{j_A}{2M_P^2} \Big]. \label{dt_lam}
\end{equation}
 The first-order form of the Klein-Gordon equation is given by
\begin{align}
    \dot{\Phi}=&\frac{\alpha}{\sqrt{A}B}\Pi,\quad\dot{\Psi}=\partial_r\left(\frac{\alpha}{\sqrt{A}B}\Pi\right),\quad\dot{\Pi}=\frac{1}{r^2}\partial_r\left(\frac{\alpha Br^2}{\sqrt{A}}\Psi\right),\label{KGE_1st}
\end{align}
The complete set of dynamical variables of this system are $\{\alpha,\;A,\;B,\;K,\;K_B,\;\lambda,\;D,\;D_B,\;\Tilde{U},\;\Phi_c,\;\Psi,\;\Pi\}$ whose evolution is dictated by the above equations of motion. 

{\bf Gauge fixing and initial conditions.} Since the system evolves to a black hole singularity with a horizon, the coordinate/gauge choice must avoid the coordinate singularities to enable us to study the evolution of the system even after the horizon is formed. A computationally economic gauge choice is Bona-Masso type slicing \cite{BonaMasso}, where the lapse is treated as a dynamical variable that obeys a wave equation sourced by $K$,
\begin{align}
    \dot{\alpha}=-\alpha^2f(\alpha)K,\quad\dot{D}_\alpha=-\partial_r(\alpha f(\alpha)K).
\end{align}
Here $f(\alpha)$ is a positive but arbitrary function of $\alpha$, which should be chosen judiciously according to the problem. For critical collapse scenarios, $1+\log$ gauge choice is preferred for which $f(\alpha)=2/\alpha$ \cite{alcubierre}. 

The goal is to discretize the EKG equations on the grid and to evolve the system of partial differential equations numerically. The initial conditions for the geometric and the matter degrees of freedom are as follows,
\begin{align}
    K|_{t=0}=K_B|_{t=0}=0&,\quad\alpha|_{t=0}=B|_{t=0}=1,\quad D_\alpha|_{t=0}=D_B|_{t=0}=0,\\
    \Phi_c|_{t=0}=f(r)&,\quad \Psi|_{t=0}=\partial_rf(r),\quad\Pi|_{t=0}=0.
\end{align}
The metric variable $A(r)$ at $t=0$ is obtained by integrating the Hamiltonian constraint on the initial hypersurface with $A(r=0)=1$
\begin{align}
    \partial_rA^0=A^0\left(\frac{1-A^0}{r}+\frac{r}{2M_P^2}(\Psi^0)^2\right).
\end{align}
This initial data is evolved using the Einstein equations, and at each iteration the $L^2-$norm of the Hamiltonian constraint is evaluated. We consider the Gaussian initial profile for the background scalar field,
\begin{align}
    f(r)=A\;exp\left[-\frac{r^2}{D^2}\right],\label{Profile1}
\end{align}
which is parameterized by $A~\text{and}\;D$. The classical analysis of this system yields a family of black hole solutions (only for a particular range of values of amplitude $A$) where the amplitude is the preferred parameter. The analysis by Berczi et. al. \cite{Berczi_PRD,Berczi_JHEP,Berczi:2024yhb} showed that the numerical evolution for the semiclassical system forms the black holes, and the solutions follow the same critical behavior as predicted by Choptuik \cite{Choptuik} with the critical value of the amplitude being $A_{crit}=1.87$ for $D=1$. We study the quantum correlations for the test field propagating on a collapsing geometry, that is, a classical field is sourcing the collapse, and the propagating quantum field on the classical geometry has no backreaction on the background dynamics. In the next subsection, we discuss the quantization of the scalar field in this background geometry, which will be denoted by the symbol $\hat{\Phi}_{q}$  to distinguish it from the collapsing classical field. 



\subsection{Test field dynamics}

In spherically symmetric spacetimes, the test quantum field can be expanded as
\begin{align}
    \hat{\Phi}_q(t,r)=\sum_{l,m}\int dk\left[\hat{a}_{k,l,m}\Tilde{u}_{k,l}Y_l^m(\theta,\phi)+\hat{a}^\dagger_{k,l,m}\Tilde{u}^*_{k,l}Y_l^{m*}(\theta,\phi)\right].
\end{align}
Here $Y_l^m$ are the spherical harmonics and $\Tilde{u}_{k,l}$ are the rescaled mode functions $\Tilde{u}_{k,l}=u_{k,l}/r^l$. The Fock space of the quantum field is built using the ladder operators $\hat{a}_{k,l,m}$ that satisfy the commutation relation $[\hat{a}_{k,l,m},\hat{a}^\dagger_{k',l',m'}]=\;\delta_{ll'}\delta_{mm'}\delta(k-k')$. The vacuum of the theory is defined as $\hat{a}_{k,l,m}\ket{0}=0,\;\forall\;k,l,m$. 

The vacuum correlations of the test quantum field in terms of mode functions take the form
\begin{align}
    \langle 0| \hat{\Phi}_q(t,r') \hat{\Phi}_q(t,r) | 0\rangle = \frac{\hbar c^2}{4 \pi} \int \, \sum_{l}^{} (2l+1) \Tilde{u}_{k,l}(t,r')\Tilde{u}_{k,l}^*(t,r)dk.
\end{align}
The mode sum representation of the two point function of the test field is divergent and needs to be regularized. Since the divergences in the correlations are directly related to the local curvature of spacetime, the regularization scheme involves adding counter terms of geometrical origin which is a challenging task on a lattice. We follow the prescription introduced in \cite{Berczi_PRD} where the expectation value of the stress-energy tensor is regularized using Pauli-Villars regularization \cite{Pauli:1949zm,Visser:2016mtr,Kamenshchik:2018ttr}. In this approach, auxiliary massive quantum fields are introduced that cancel the divergent contributions to the stress-energy tensor. A proper regularization requires the addition of $5$ massive quantum fields that remove the divergences caused by the test field as well as the divergences introduced by the auxiliary field. The action for the test field after incorporating the regularization prescription is extended to
\begin{align}
    \mathcal{L}_{test} = -\sum_{n=0}^5(-1)^n\left[\frac{1}{2}\partial_a\Phi^n_q\partial^a\Phi^n_q+\frac{1}{2}m_n^2(\Phi_q^n)^2\right]
\end{align}
where $n=0$ corresponds to the massless test field with auxiliary fields with masses $m_1=m_3$, $m_2=m_4=\sqrt{3}m_1$ and $m_5=2m_1$. A detailed discussion on the regularization scheme can be found in \cite{Berczi_JHEP,Berczi:2024yhb}. The regularized correlation function for the test field takes the form
\begin{align}
    \langle0| \hat{\Phi}_q(t,r') \hat{\Phi}_q(t,r) | 0\rangle^{reg} = \frac{\hbar c^2}{4 \pi} \int \,\sum_{n=0}^5 \sum_{l}^{}(-1)^n (2l+1) \Tilde{u}_{k,l;n}(t,r')\Tilde{u}_{k,l;n}^*(t,r)dk.\label{PhiPhi_corr}
\end{align}
Similarly, the equal time correlation function for the momentum operator conjugate to the field variable takes the form
\begin{equation}\label{pipi_corr}
\begin{split}
    \langle0| \hat{\Pi}_q(t,r') \hat{\Pi}_q(t,r) | 0\rangle^{reg} =\frac{\hbar c^2}{4 \pi}&\frac{\sqrt{A(t,r')}B(t,r')}{\alpha(t,r')}\frac{\sqrt{A(t,r)}B(t,r)}{\alpha(t,r)} \\
    &\int \,\sum_{n=0}^5 \sum_{l}^{}(-1)^n (2l+1)\partial_t\Tilde{u}_{k,l;n}(t,r')\partial_t\Tilde{u}_{k,l;n}^*(t,r)dk.
\end{split}
\end{equation}
The dynamics of the quantum fluctuations, as well as the expectation value of the stress-energy tensor, are determined by the evolution of the mode functions. The first-order evolution equations for field modes are
\begin{equation}\label{ModeEvo}
\begin{split}
    \dot{u}_{k,l;n}=&\frac{\alpha}{\sqrt{A}B}\pi_{k,l;n},\quad\dot{\psi}_{k,l;n}=\partial_r\left(\frac{\alpha}{\sqrt{A}B}\pi_{k,l;n}\right),\\
    \dot{\pi}_{k,l;n}=&\partial_r\left(\frac{\alpha}{\sqrt{A}B}\right)\left(\frac{l}{r}u_{k,l;n}+\psi_{k,l;n}\right)+\frac{\alpha\beta}{\sqrt{A}}\left(\frac{2l+2}{r}\psi_{k,l;n}+\partial_r\psi_{k,l;n}\right)\\
    &~~~~+\frac{l(l+1)}{r^2}\left(\frac{B}{A}-1\right)\sqrt{A}\alpha u_{k,l;n}-m_n^2\alpha\sqrt{A}Bu_{kl;n},
\end{split}
\end{equation}
where $\psi_{k,l;n}=\partial_ru_{k,l;n}$, which together with \cref{dt_AB,dt_DBl,dt_Utilda,dt_K,dt_KB,dt_lam,KGE_1st} form the complete set of coupled differential equations that determine the dynamics of the background variables and the test field.  

The initial conditions for the mode functions are obtained by identifying their spatial profile at the initial hypersurface with the mode function in the Minkowski spacetime given in terms of spherical Bessel functions
\begin{align}
    u^0_{k,l;n}(0,r)=\frac{k}{\sqrt{\pi\omega_n}}\frac{j_l(kr)}{r^l},\quad\psi^0_{k,l;n}(0,r)=\frac{k}{\sqrt{\pi\omega_n}}\left(\frac{j'_l(kr)}{r^l}-\frac{lj_l(kr)}{r^{l+1}}\right),\quad \pi^0_{k,l;n}(0,r)=-\frac{i\omega_n k}{\sqrt{\pi\omega_n}}\frac{j_l(kr)}{r^l},
\end{align}
where $\omega_n=\sqrt{k^2+m_n^2}$. With these initial conditions, the mode functions are evolved using their equations of motion \cref{ModeEvo} together with the EKG system. For the numerical implementation of these coupled partial differential equations for the background variables and the test field, we modified the code developed by Berczi et al. \cite{Berczi_JHEP,BercziGH} for the collapse of the quantum field to our set up, using the discretization and optimization framework they introduced. In the next subsection, we briefly discuss these discretization and optimization schemes.

\subsection{Methodology}
The non-linear, coupled equations presented in previous subsections are solve through numerical methods by discretizing the first-order equations on a lattice. We implement a uniform spatial grid comprising 500 points with a grid spacing of dr = 0.025 and a time step of $\displaystyle dt = dr/4$. We employ tenth-order finite difference methods to compute spatial derivatives, while the time integration utilizes the tenth-order Runge-Kutta (RK) method. The evolution equations that we are working with include terms such as $\displaystyle r^{-1}$ and $\displaystyle r^{-2}$ that give rise to numerical instability near the origin. To mitigate or postpone the emergence of errors, artificial dissipation is introduced, which functions as a damping mechanism for those errors. We incorporate Kreiss-Oliger terms \cite{kreiss1973methods} into the evolution equation for each field. For example, for $\Phi_c (t,r)$ at the time step $n$, with the original evolution step represented schematically by $G(\Phi_c^n)$, the value of $\Phi_c$ at the subsequent time step is determined by: 
\begin{align}
    \Phi_c^{n+1} = \Phi_c^n + G(\Phi_c^n)dt
\end{align}
With dissipation, the evolution equation G($\Phi_c^n$) is modified as: 
\begin{align}
    G' (\Phi_c^n) = G(\Phi_c^n) - \epsilon (-1)^N dr^{2N-1} \partial_r^{2N} \Phi_c^n 
\end{align}
where $\epsilon $ is a positive constant, and 2N is the order of dissipation. This added term damps the modes with a wavelength close to the grid spacing $dr$. The Kreiss-Oliger terms as 4th order with $\epsilon =0.5$ for the quantum mode functions and $\epsilon = 0.1$ for the metric fields and classical scalar field give better stability of the numerical evolution. The mass $m_1$  of the ghost field $G_1$ is taken to be 1 in Planck units throughout our simulations.  The appearance of the continuous integrals of the mode functions over the wave number $k$ in the equal time correlation functions of the test quantum field needs to be discretized as well. To do this, a minimum value of the wave number $k_{min}$ is identified that depends on the size of the radial box, which will then be the minimal step $dk = k_{min}$ that appear in the mode sum representation of the integrals over $k$. After careful consideration of stability, we choose $\displaystyle k_{min} = \pi/15$. 

Besides the Gaussian parameters, the most sensitive part of the simulation are the quantum mode functions with high $l$ values that often develop instabilities \cite{Berczi_JHEP,Berczi:2024yhb}. Although we have added proper dissipation, the instabilities still arise and limit the number of $l$ modes that can be added, as dissipation does not completely remove the instability but delays it. If we run the simulation for a long time, or use extremely high $l$, we are bound to get unstable evolution. Thus, the choices for the parameters used in the simulation ($A$, $D$, and the number of $k$ and $l$ values) can be varied only in a particular domain. The simulation breaks down outside that domain leading to violation of the constraints and unstable dynamics of the mode functions. We take $N_l = N_k = 50$ to ensure that the mode functions are stable in the spatial grid of the size that can enable us to do a meaningful analysis of the field correlations inside as well as outside the black hole horizon. This value could be varied depending on the range of the spatiotemporal zone, the type of dissipation added, and the profile of the test field itself, for more details, see \cite{Berczi_JHEP,Berczi:2024yhb}. In our analysis, we have not considered near-critical scenarios as in this case the horizon formation takes a long time and develops numerical instabilities. We focus on the subcritical and supercritical scenarios, where we take the amplitude of the background scalar field profile to be $A=1<A_{crit}$ and $A=5>A_{crit}$.

\section{Gravitational collapse: Background dynamics}\label{GC_BD}

\begin{figure}
    \centering
    \includegraphics[width=0.99\linewidth]{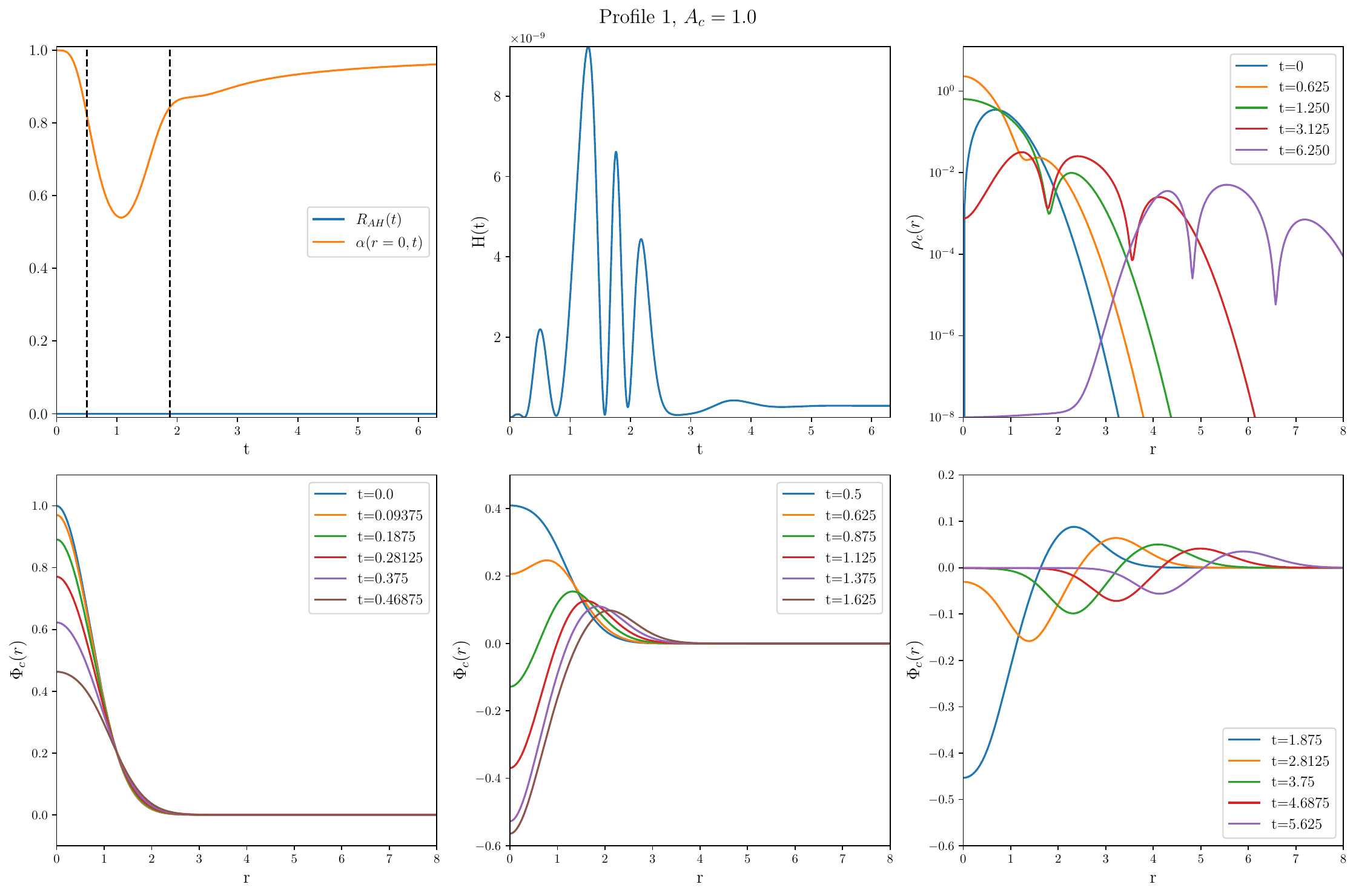}
    \caption{Subcritical evolution of background observables. In the first frame of the first row, we have the time evolution of the lapse function at the origin (orange curve) and the areal radius of the apparent horizon (blue curve). The second frame depicts the time evolution of the Hamiltonian constraint. In the third frame of the first row, we have the spatial profile of the energy density of the background scalar field at different instances during the evolution. In the second row, we show the time evolution of the spatial profile of the background scalar field in three regimes marked by the dashed black lines in the first frame of the first row. In the first frame, we consider the instance during which the maximum of the profile at the origin is decreasing. In the second frame, the nature of extrema at the origin changes with a global maximum away from the origin. The last frame illustrates a pulse of a scalar field evolving away from the origin towards the asymptotic observer.}
    \label{fig:bg_dyn_P1_NBH}
\end{figure}

\begin{figure}
    \centering
    \includegraphics[width=0.99\linewidth]{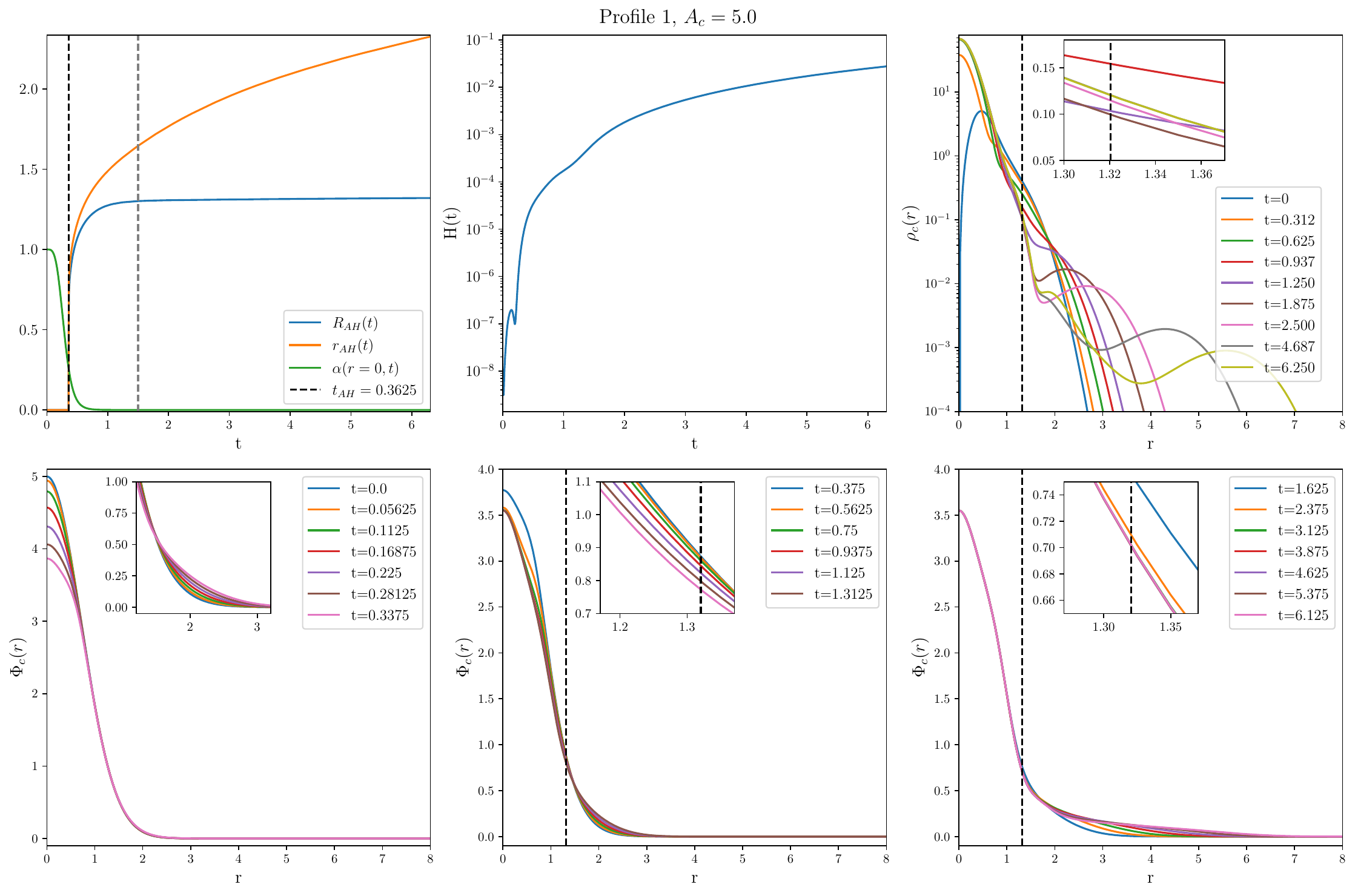}
    \caption{Supercritical evolution of background observables. In the first frame of the first row, we have the time evolution of the lapse function at the origin (green curve), the areal radius of the apparent horizon (blue curve), and the location of the apparent horizon (orange curve). The vertical dashed black line marks the instance when the horizon appears, and we consider the horizon evolving for the time between the dashed black line and the dashed gray line. The second frame depicts the time evolution of the $L^2$ norm of the Hamiltonian constraint. In the third frame of the first row, we have the spatial profile of the energy density of the background scalar field at different instances during the evolution. The inset figure shows the near-horizon behavior of the energy density. In the second row, we show the time evolution of the spatial profile of the background scalar field. We consider the instance before the horizon appears in the first frame, the period for which the horizon is evolving is considered in the second frame, and in the last frame, the areal radius of the horizon is no longer changing. The vertical dashed lines represent the areal radius of the apparent horizon.}
    \label{fig:bg_dyn_P1_BH}
\end{figure}

We now present the dynamics of the background geometry for the two cases (i) subcritical (no black hole, flat spacetime), and (ii) supercritical (black hole spacetime) evolution arising from the initial conditions and the scalar field profile \cref{Profile1}. The stability of the simulation setup for the background evolution is discussed in detail in \cite{Berczi_JHEP}, and we only discuss the behavior of the $L^2-$norm of Hamiltonian constraint as the indicator of the accuracy of the simulation. The numerical evolution of the system is expected to break when the $L^2-$norm of Hamiltonian constraint is of $O(1)$. For the gravitational sector, we look at the evolution of the lapse function at the origin which vanishes when the black hole forms with the coordinate choice. The location of the apparent horizon $r_{AH}$ is determined by checking the outermost radius where the expansion of null rays
\begin{align}
    \Theta = \frac{1}{\sqrt{A}}\left(\frac{2}{r}+\frac{\partial_rB}{B}\right)-2K_B\label{expansion}
\end{align}
vanishes. The areal radius of the apparent horizon is $R_{AH}=r_{AH}\sqrt{B|_{r_{AH}}}$ which gives the physical distance of the two-dimensional hypersurface located at $r=r_{AH}$ from the origin. For the matter sector, we are interested in the evolution of the energy density and spatial profile of the background scalar field. The results of the background evolution for the initial data that leads to subcritical evolution are presented in \cref{fig:bg_dyn_P1_NBH}, and the supercritical evolution is shown in \cref{fig:bg_dyn_P1_BH}. 

In the first frame of the first row of \cref{fig:bg_dyn_P1_NBH}, we have the evolution of the lapse function at origin $\alpha(0,t)$ (orange curve), which starts from unity and decreases to attain a minimum and starts increasing towards unity. Since the expansion of null rays and lapse at the origin never vanishes, the gravitational evolution does not lead to the formation of black hole. The second frame shows the time evolution of the $L^2-$norm of the Hamiltonian constraint, which has successive peaks around the time when the lapse function has its minimum, although the constraint violation is still small in this subcritical evolution. The third frame shows the spatial profile of the energy density of the background scalar field at different instances during the evolution. Initially, the energy density is a single peak profile with a global maximum at finite radial coordinate. As the system evolves, the peak moves toward the origin and reflects and scatter to infinity. This behavior is reflected by the evolution of the spatial profile of the background scalar field presented in the second row of \cref{fig:bg_dyn_P1_NBH}. We separate the scalar field dynamics in three regimes marked by vertical dashed lines in the first frame of the first row, at first the location of the global maximum is at the origin and its magnitude decreasing, then the nature of extrema changes at the origin, and finally a pulse of maximum and minimum forms and evolve outwards. 

\begin{wrapfigure}{r}{0.5\textwidth}
    \centering
    \includegraphics[width=0.99\linewidth]{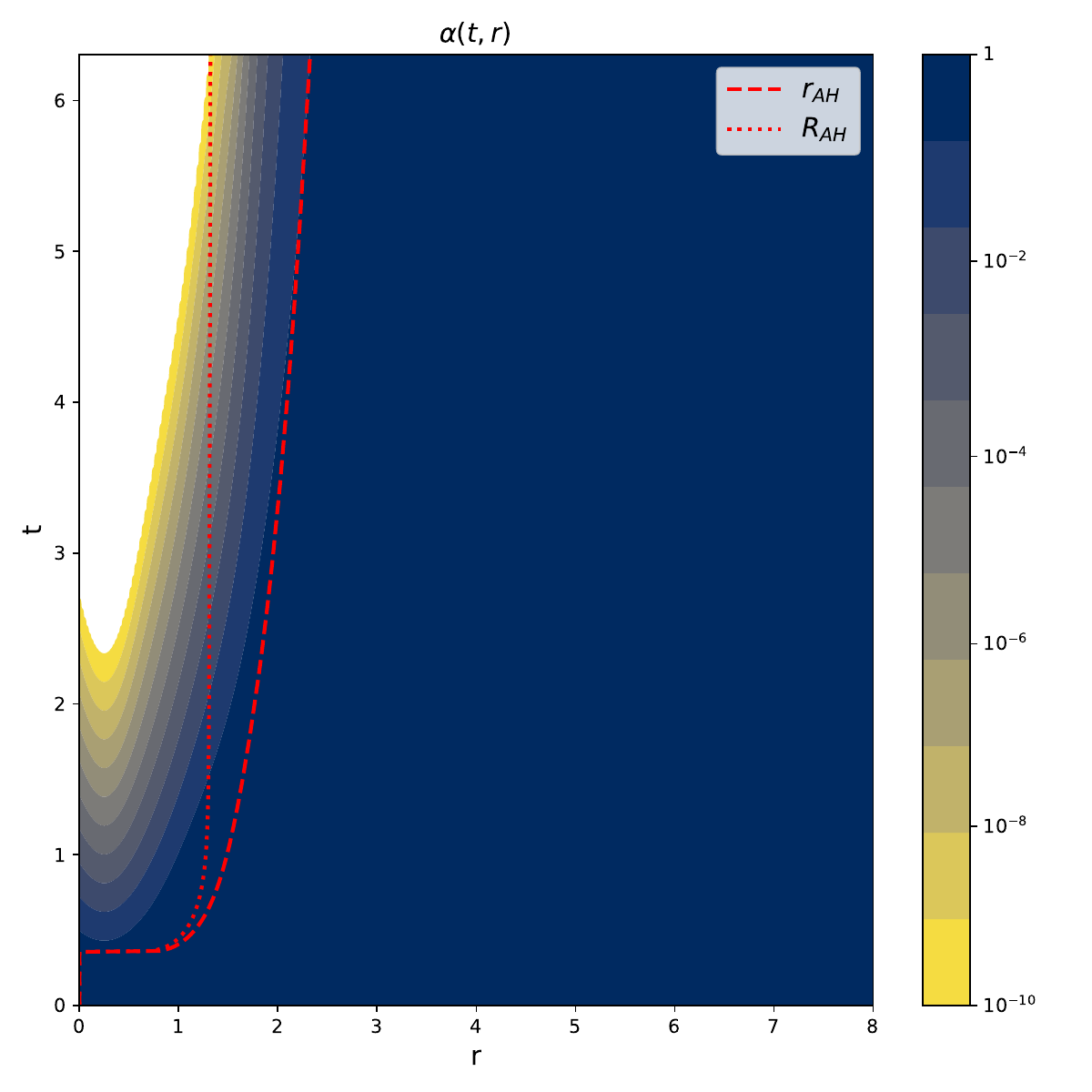}
    \caption{Time evolution of lapse profile in the supercritical scenario depicting the collapse of the lapse. The dotted line represents the areal radius, and the dashed line corresponds to the location of the apparent horizon.}
    \label{fig:lapse}
\end{wrapfigure}

In \cref{fig:bg_dyn_P1_BH}, we present the background dynamics for the case when the black hole forms. In the first frame, we plot the lapse function at the origin (green curve) along with the areal radius of the apparent horizon $R_{AH}$ (blue curve) and the location of the apparent horizon $r_{AH}$ (orange curve) as a function of time. The apparent horizon appears at time $t_{AH}=0.3625$ with areal radius $R_{AH}=0.725$ marked by a vertical black dashed line, and it grows sharply to a value $R_{AH}=1.30$ by $t=1.5$ and then rises slowly to the areal radius of order $R_{AH}=1.32$ by the end of the simulation. We will focus on these three regimes, namely no horizon, apparent horizon, and stationary horizon to investigate both the dynamics of the background scalar field and the test quantum field. The location of the apparent horizon continues to grow steadily even during the stationary phase. The lapse at origin is small but non-zero during the apparent horizon regime, and then it decays exponentially during the stationary horizon phase. Away from the origin, the lapse is practically zero for $r<R_{AH}$ but is small but finite for $R_{AH}<r<r_{AH}$ during the stationary horizon phase, as seen in the \cref{fig:lapse}. The second frame in the first row shows the evolution of the Hamiltonian constraint, where we see a minimal constraint violation for most of the simulation duration. At the later stages, the $L^2-$norm of the Hamiltonian constraint rises steadily to take the maximum value $0.02$ at the end of the simulation, effectively a constraint preserving evolution as the $L^2-$norm is negligible as compared to other energy scales natural to the setup. 

The third frame of \cref{fig:bg_dyn_P1_BH} shows the time evolution of the spatial profile of the energy density of the background scalar field during different phases of evolution. We see that the scalar field energy density initially peaked at a finite radius and moved towards the origin. Instead of bouncing back, the energy density of the scalar field freezes near the origin as soon as the horizon appears due to the {\it collapse of the lapse}, with a profile of a maximum central density at the origin and exponentially decaying behavior. As the system evolves during the apparent horizon phase, the energy density flattens around the areal radius of the stationary horizon with significant changes near the horizon for $R_{AH}<r<r_{AH}$. Once the evolution enters the stationary horizon phase, the energy density profile inside the horizon is stationary, with only minor changes occurring near the horizon, as illustrated in the inset figure. However, the story outside the horizon is intriguing. As the horizon is still expanding, the field receives a final burst of energy, which manifests as a bump around $t\sim1.25$, which travels away from the horizon with a decaying amplitude. 

This behavior can be further understood from the dynamics of the radial profile of the background scalar field, as shown in the second row of \cref{fig:bg_dyn_P1_BH}. Initially, during the no horizon phase, the scalar fields behave the same way as the subcritical case, where only its central maximum decreases with small changes at the knee of the profile, as shown in the inset figure of the first frame. During the apparent horizon phase, the central maximum decreases further to arrive at the stationary value at the origin, and inside the horizon, the profiles change marginally. The scalar field outside receives the push during this phase, which generates the gradient in the field that travels outwards in the form of energy density during the stationary horizon phase. On the other hand, the scalar field profile inside the horizon freezes, and the dynamics is only in the exterior region in the final phase. This freezing of physical observables results from the gauge choice, where the lapse function vanishes inside the horizon for $r<R_{AH}$, leading to no time evolution in this domain and marginal changes near the horizon with $R_{AH}<r<r_{AH}$. With this general idea of the background evolution, we present the dynamics of the quantum correlations of the test field in the next section.

\section{Quantum correlations of the test field}\label{QC_TF}
\begin{figure}[!htbp]
    \centering
    \includegraphics[width=0.85\linewidth]{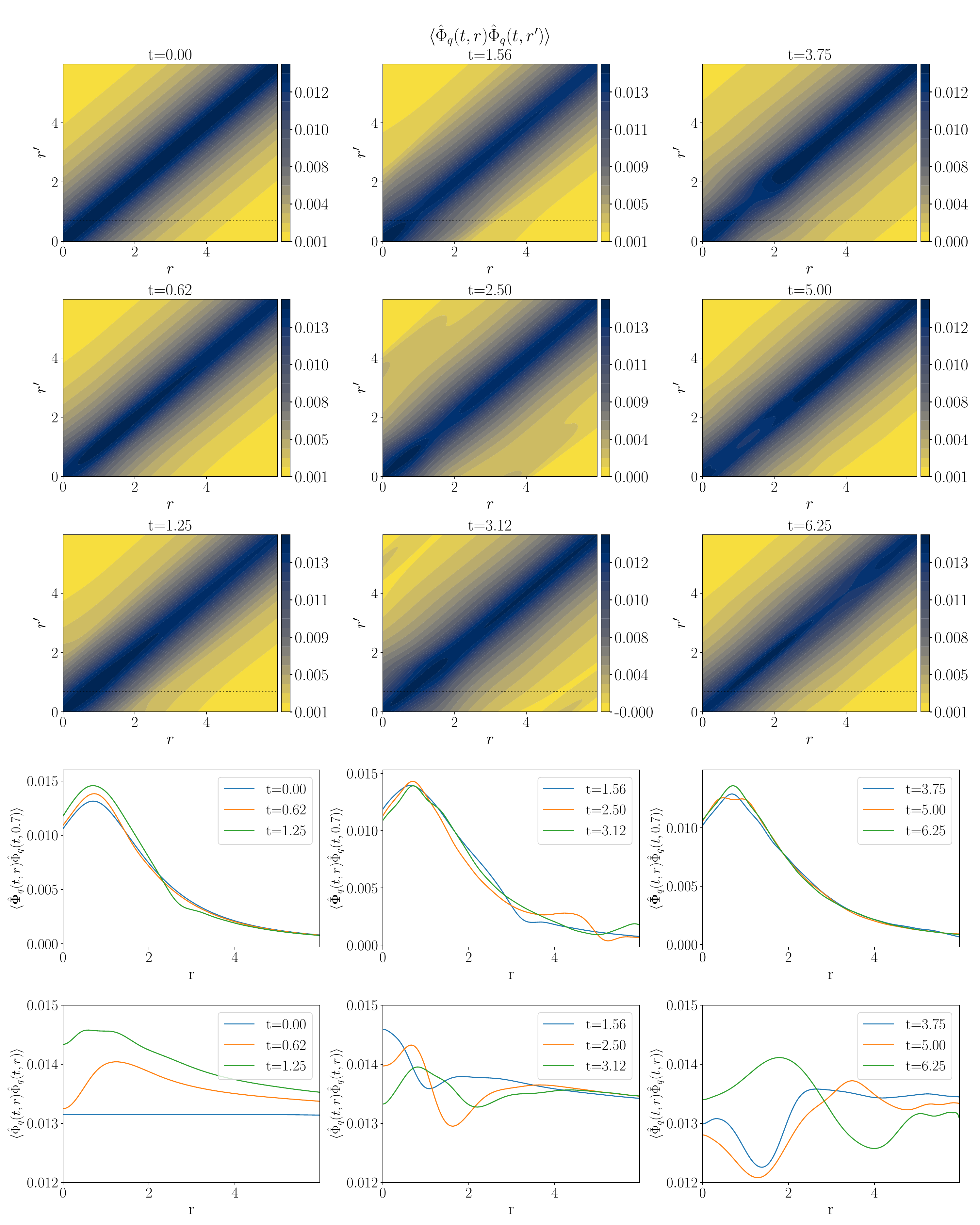}
    \caption{Evolution of the equal time correlations of the test quantum field for parameter choices that lead to subcritical evolution. We are considering the three regimes of the background dynamics for the subcritical evolution discussed in the previous section in the first three rows. The fourth row show the spatial profile of the quantum correlation along the horizontal cut represented by the black dotted line in the contour plots at the instances of the preceding frames in the corresponding column. The last row depicts the dynamics of the quantum correlation along the diagonal.}
    \label{fig:QC_P1_NBH}
\end{figure}
We are interested in the vacuum correlations of the test quantum field and its conjugate momentum given in \cref{PhiPhi_corr,pipi_corr} in the subcritical and supercritical scenarios with initial profile given in \cref{Profile1}\footnote{The generic features of the quantum correlations that we have presented for the Gaussian initial data in \cref{Profile1} are observed for other initial profiles as well. We checked it for profile
\begin{align}
    f(r)=A\left(\frac{r}{D}\right)^2\left[exp\left(-\left(\frac{r-R}{D}\right)^2\right)+exp\left(-\left(\frac{r+R}{D}\right)^2\right)\right],\label{Profile2}
\end{align}
though results are not presented in the manuscript.}, that we discussed in the previous section. As alluded to in the Introduction, the motivation to study the quantum correlations is multifold; the first is related to Hawking radiation and black hole atmosphere, while the other being the observability of Hawking radiation through analogue black hole experiments. The standard understanding of the Hawking process is that the Hawking pair is created near the horizon \cite{Hawking_1976,Hotta:2015yla}, where one particle moves inward towards the central singularity and the other escapes and moves towards the asymptotic observer. This is expected to translates into non-local correlations between the quantum field excitation inside and outside the horizon \cite{Schutzhold:2010ig}. The presence of such correlations indicates not only the presence of Hawking radiation but also their characteristics, further indicating where these excitations originate \cite{Unruh_atmosphere,Schutzhold:2008tx,Schutzhold:2010ig,Fontana:2023zqz,Giddings:2015uzr,Balbinot:2021bnp,Kaczmarek:2023kpn}. Another aspect that we intend to investigate is the correspondence between the analogue signatures of Hawking radiation through the density correlations \cite{Balbinot,Carusotto:2008ep} and the quantum correlation in collapsing geometry \cite{Balbinot:2021bnp,Schutzhold:2010ig}. A related analysis appeared recently that investigated the quantum correlations of the field that sources the geometry \cite{Berczi:2024yhb}, however a more accurate correspondence with the Hawking process would be a test quantum field propagating on the dynamical black hole spacetime, which we have considered. 

The time evolution of the quantum correlations for the subcritical scenario is presented in \cref{fig:QC_P1_NBH}, where we have the heatmap of the equal time correlations at different stages of evolution in the first three rows. The next row shows the spatial profile of the correlations along the dashed horizontal line at the time considered in the preceding frames. In the last row, we plot the spatial profile of the correlations along the diagonal at different stages of the evolution. At the beginning of the simulation, the correlations have a constant maximum along the diagonal, with off-diagonal correlations decaying exponentially. A global maximum for the correlations appears along the diagonal, and its magnitude, as well as the correlations at the origin, increases as the system evolves further. This peak in correlations moves towards the origin and is then reflected from the origin, leading to oscillatory features appearing afterward, as seen in the last row in \cref{fig:QC_P1_NBH}. The non-local behavior of the correlations can be seen from their profile along the horizontal cut, where the correlations first increase around $r=r'$ and then are propagated away as small oscillations near the tail.

\begin{figure}
    \centering
    \includegraphics[width=0.85\linewidth]{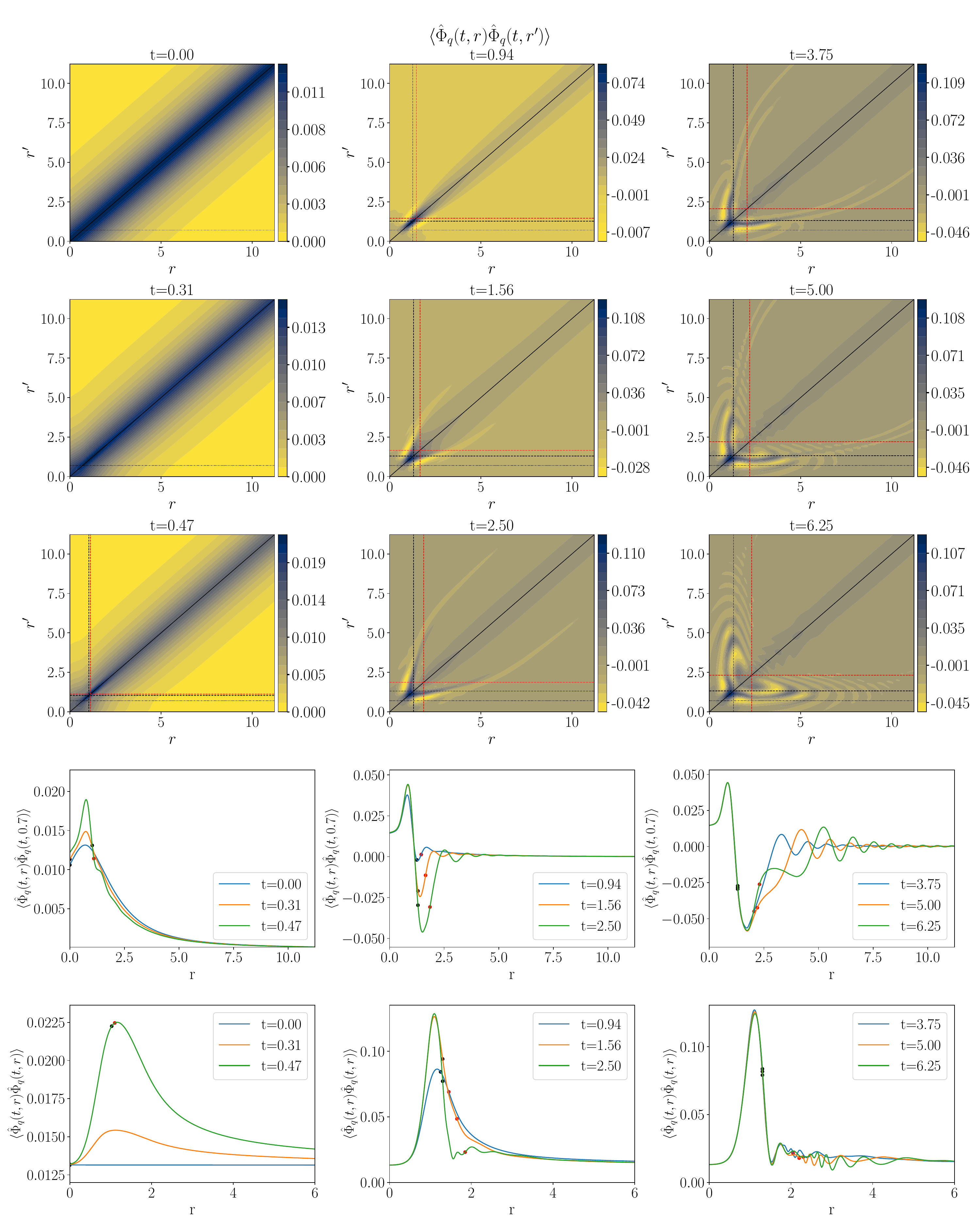}
    \caption{Evolution of the equal time correlations of the test quantum field for parameter choices that evolves to black hole background. We are considering the three regimes of the background dynamics for the supercritical scenario and present the evolution of vacuum correlations of the test field through the contour plots in the first three rows. The dashed black line is the areal radius of the apparent horizon $R_{AH}$, and the dashed red line is the location of the apparent horizon $r_{AH}$. The plots in the fourth row depict the behavior of the correlations along the horizontal cut inside the apparent horizon, with a black dot representing the areal radius of the horizon and a red dot representing the location of the horizon. The last row depicts the evolution of the vacuum correlations along the diagonal $(r=r')$ with black and red dots representing the areal radius and location of the horizon, respectively.}
    \label{fig:QC_P1_BH}
\end{figure}

On the other hand, the vacuum correlations show interesting behavior in the supercritical scenario. We are following the same layout in \cref{fig:QC_P1_BH} as previously, where the first three rows shows the heatmap for the equal time correlations at different stages of the evolution with the next row showing the behavior of correlations at fixed $r'$ inside the horizon and the last row depicts the local behavior $(r=r')$ of the correlations. Before the horizon appears, the constant maximum along the diagonal localizes and follows the dynamics akin to the subcritical dynamics. Once the horizon forms, the location of the global maximum along the diagonal freezes, but the magnitude keeps on increasing, and the width keeps on decreasing during the apparent horizon regime. Once the evolution enters the stationary horizon phase, the correlations decrease sharply inside the horizon for $r>R_{AH}$ and attain a local minimum just inside the horizon for $R_{AH}<r<r_{AH}$. Outside the horizon, oscillations originate in the local correlations that propagate away from the horizon as the system evolve, as seen in the last row of \cref{fig:QC_P1_BH}. Moreover, the correlations along the diagonal $r=r'$ maintain their amplitude to the same order as their initial constant value at $t=0$ throughout the evolution. Even with the buildup of non-local features in stationary horizon phase, correlations in the patch around the diagonal are preserved outside the horizon, with the shape of this patch getting distorted. Another thing to note is that the correlations along the diagonal never become negative while the global maximum decreases slowly, as shown in the last row of \cref{fig:QC_P1_BH}, hinting at residual dynamics even in the regime where the lapse is vanishing.


Non-local features are seen to emerge away from the diagonal $(r\neq r')$ as the horizon appears. A valley of negative correlations appear parallel to the local peak on the main diagonal on both sides. During the apparent horizon regime, the strong positive correlations along the main diagonal get bifurcated in two directions, with the negative correlations also getting tilted. The positive correlations along the two prongs get elongated at first and get separated from the main diagonal, with negative correlations still confined to the outer side of each prong. As the system evolves further, negative correlations start to grow on the outer side of the fork as well, leading to the positively correlated widened fork that is enveloped by a negatively correlated patch. These non-local forks attain further oscillatory character at the same time as the oscillations appear on the main diagonal. At this stage, rather than leaning toward the origin as seen in \cite{Berczi:2024yhb}, the correlations start to tilt toward the horizon. Patches of crests and troughs begin to form along the diagonal with the shape of a crescent that originates from the diagonal and ends at the horizon. Interestingly, the prominent non-local features are masked by the apparent horizon $r=r_{AH}$ as observed in \cite{Berczi:2024yhb} with the difference being the near horizon character of the correlations.

\begin{figure}
    \centering
    \includegraphics[width=0.9\linewidth]{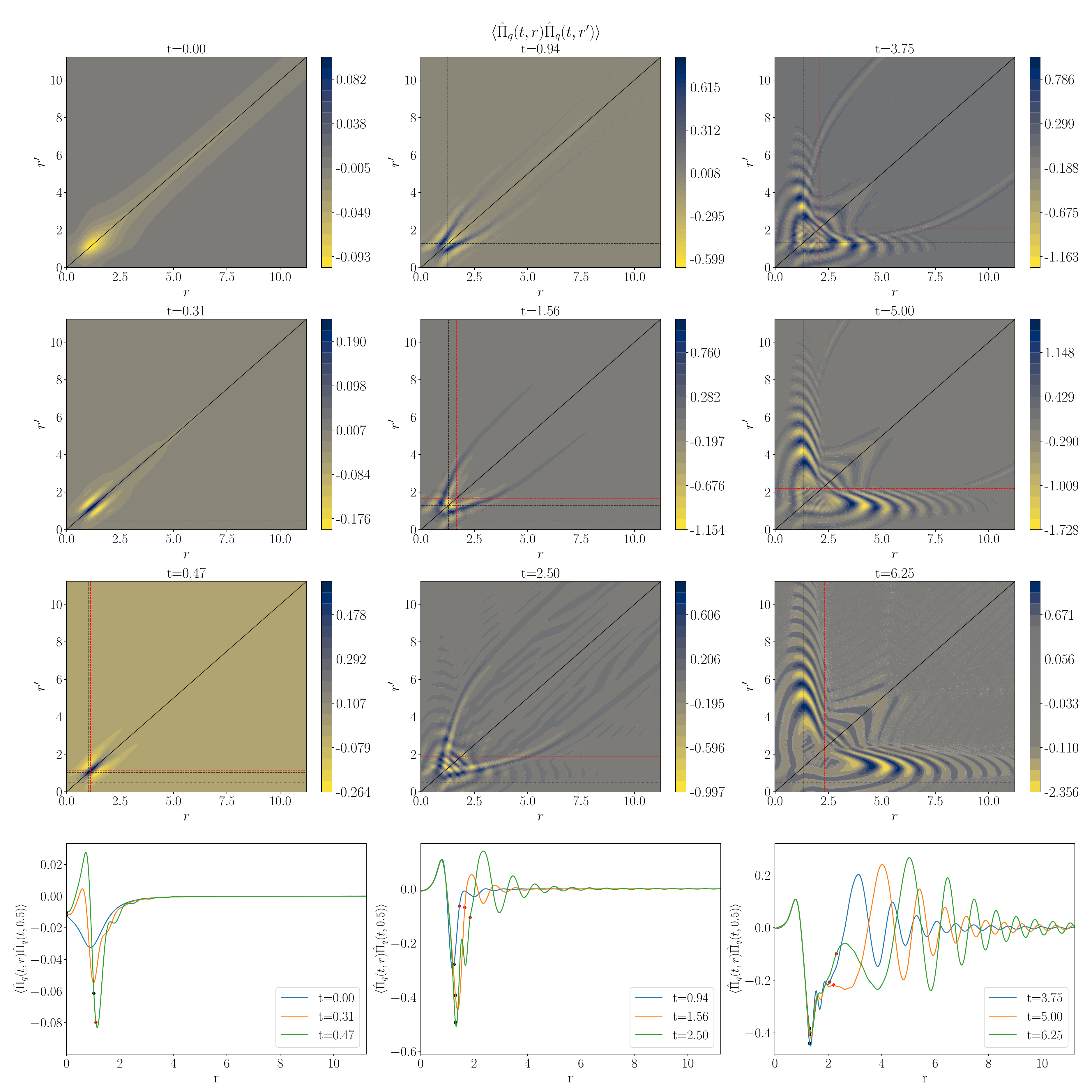}
    \caption{Evolution of the equal time correlations of the momentum operator conjugate to the test quantum field in the first three rows, for parameter choices that evolve to black hole background. The dashed black line is the areal radius of the apparent horizon $R_{AH}$, and the dashed red line is the location of the apparent horizon $r_{AH}$. The last row depicts the evolution of the momentum correlations along the horizontal cut inside the horizon, with black and red dots representing the areal radius and location of the horizon, respectively.}
    \label{fig:QMC_P1_BH}
\end{figure}

\begin{figure}
    \centering
    \begin{tabular}{cc}
       \includegraphics[width=0.38\linewidth]{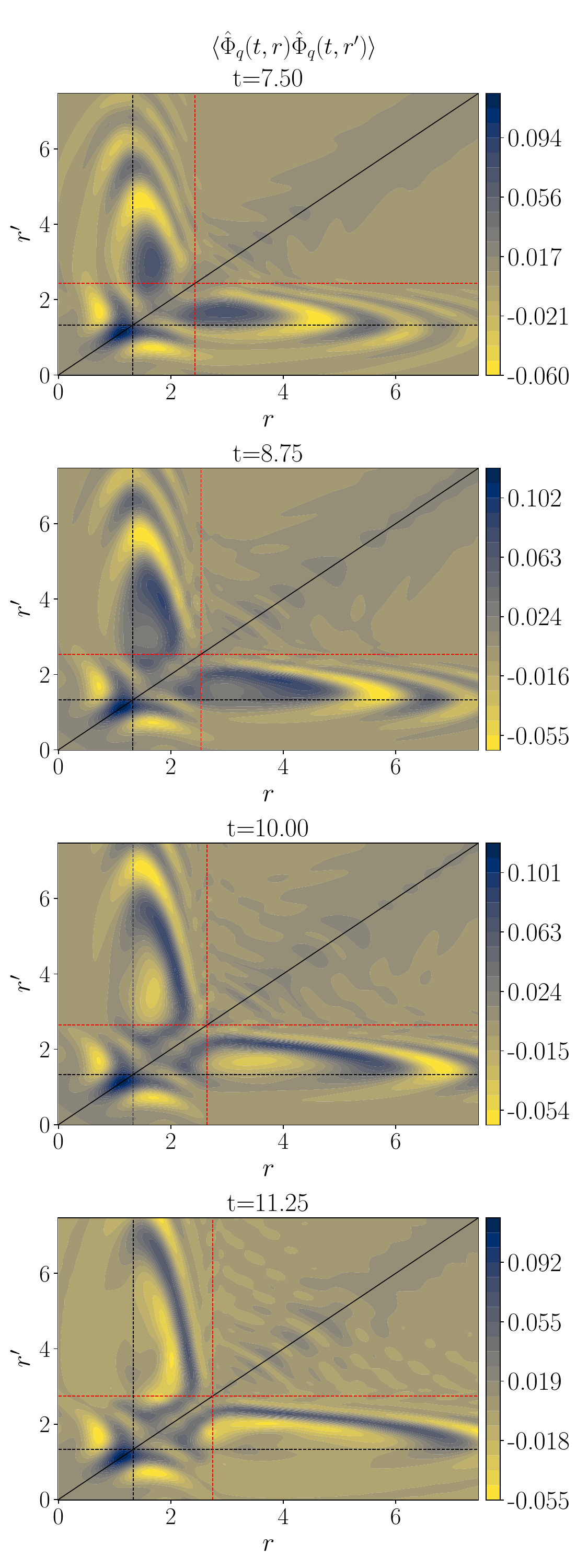} & \includegraphics[width=0.38\linewidth]{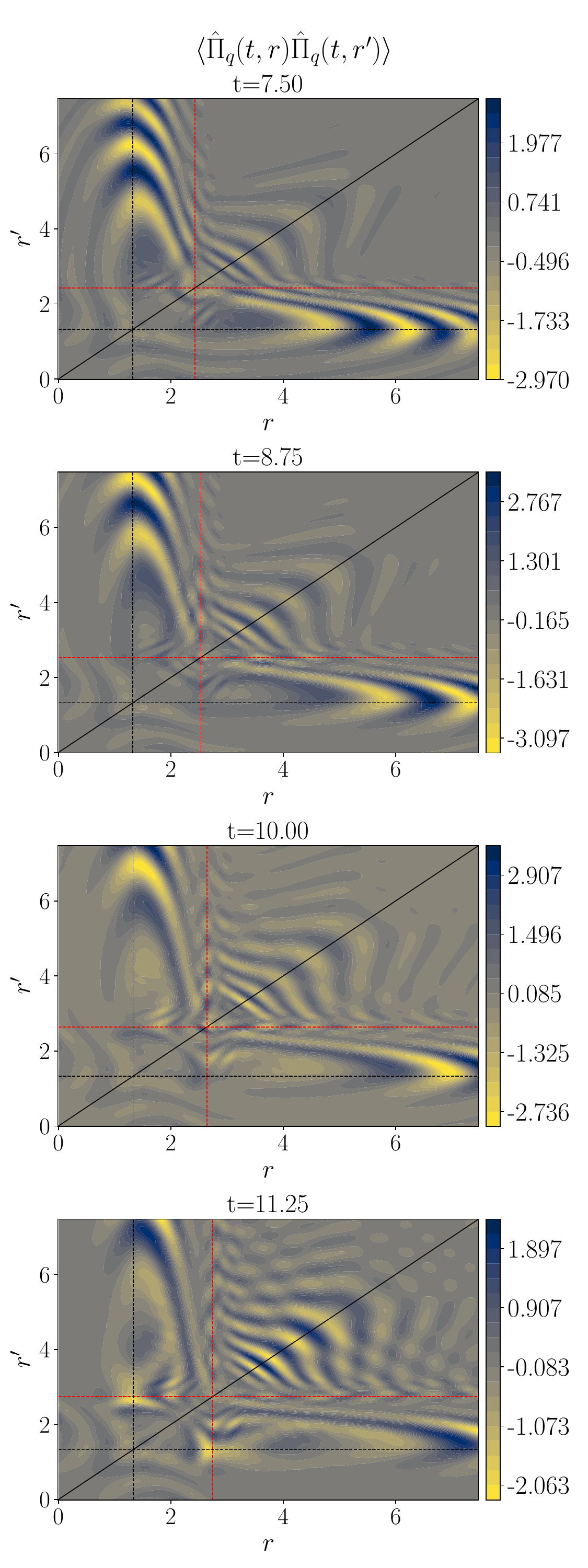}
    \end{tabular}
    \caption{Late-time evolution of the equal time correlations of the field operator in the first column and the momentum operator in the second column. The dashed black line is the areal radius of the apparent horizon $R_{AH}$, and the dashed red line is the location of the apparent horizon $r_{AH}$.}
    \label{fig:QMC_P1_BH_LT}
\end{figure}

The field correlations throughout the evolution retain the mirror symmetry with respect to the reflection along the diagonal. All local maxima and minima on the one side of the diagonal have their counterparts at the same distance on the line perpendicular to the diagonal. The non-local character is even stronger for the correlation function of the momentum operator $\hat{\Pi}(r,t)$ shown in \cref{fig:QMC_P1_BH}. Initially, the correlations are negative along the diagonal and it is peaked at the radius where the metric function $A$ is peaked. In this case, the oscillatory features far away from the diagonal are of the same nature as for the field correlations, which are again masked by the apparent horizon at the later states. The major difference is in the non-local behavior of the momentum correlations just outside the horizon and in the vicinity of the diagonal, where the fringe structure appears of the positive and negative correlations perpendicular to the main diagonal. The nonlocal features in the correlations get stretched as the system evolves further in the stationary regime and correlation islands of maxima and minima are localizing inside the region $R_{AH}<r<r_{AH}$ by the end of the simulation, as depicted in \cref{fig:QMC_P1_BH_LT}.

The non-local character of the field correlations is understood further by looking at the dynamics of the correlations along the horizontal cut inside the horizon $r'=r_{in}<r_{AH}$, which tells us how the field excitation inside the horizon is correlated to the excitation outside the horizon, as seen in the first three frames in the fourth row of \cref{fig:QC_P1_BH}. During the initial phase, the local correlations peak at $r=r_{in}$ increase as the system evolves. During the apparent horizon phase, negative correlations appear for $r\sim R_{AH}$ having global minimum when $R_{AH}<r<r_{AH}$ and a correlation peak outside the horizon $r>r_{AH}$ with small oscillations accompanying it. The magnitude of this peak keeps on growing during this phase as the valley of negative correlations deepens. Both global maximum as well as the valley of negative correlations settles during the stationary horizon phase. Outside the horizon, the oscillations get amplified and start traveling away from the horizon without any significant change in the amplitude. If we take the point near the apparent horizon $r'\sim r_{AH}$ or outside the horizon $r'>r_{AH}$, the correlations have a strong oscillatory behavior in addition to the local peak at $r=r'$ and this behavior is indiscernible. Similar non-local behavior is observed in the equal time correlations of field momentum operator as depicted in the last row of \cref{fig:QMC_P1_BH}.

The positive correlation peaks inside and outside the horizons are interpreted as entangled pairs of Hawking quanta \cite{Schutzhold:2010ig}. In this light, \emph{the Hawking excitation at $r'=r_{in}$ gets entangled with the excitation at $r=r_{out}$ as the horizon forms during the collapse, and the location of excitation moves away from the horizon as the system evolves}. Moreover, the correlation peak outside the horizon first appears at macroscopic distance from the location of the apparent horizon indicating that the Hawking radiation is not originating from the horizon but from the quantum atmosphere outside the horizon. Furthermore, these features are similar in spirit to the case of analogue black holes \cite{Carusotto:2008ep} where non-local correlations appear uniformly along a straight line perpendicular to the diagonal that passes through $x=0=x'$, the location of the sonic horizon. This feature of density-density correlations in the analogue black holes is manifested as a {\it characteristic peak}, which is attributed to the correlation between the Hawking particle and its entangled partner. The prediction of this peak has been the cornerstone of the analogue modeling of black hole physics, which has been observed experimentally \cite{Steinhauer:2015saa,MunozdeNova:2018fxv,Kolobov:2019qfs}. Therefore, the off-diagonal behavior of the equal time correlation function depicts a nontrivial correlation of the excitation inside the horizon with the excitation outside the horizon with the location of the outside peak moving away from the horizon towards the asymptotic observer as the system evolves as can be seen in \cref{fig:QMC_P1_BH_LT}.


\section{Conclusion}\label{Conclusion}


The Black hole information paradox has been an unresolved and a contentious issue concerning the fundamental physics. It is an area that lies at the intersection of quantum foundations, quantum information, gravitation, and (standard model) matter fields. It is, therefore, not surprising to expect its resolution as one of the guiding posts towards modeling and understanding the fundamental physics, or quantum gravity. This work is one step towards that direction. 

Within our viewpoint, the semi-classical calculation by Hawking and several others after that, calls for a full analysis wherein a dynamical gravitational (collapse) background and quantum matter fields propagating on it should be evolved self-consistently. Such an analysis can only be carried out numerically owing to the complicated system of equations and the initial value problem. Even then, the numerics is fraught with instabilities and requires a careful analysis. While gravitational collapse has been studied classically for various cases, gravitational collapse with a quantum matter field has only been studied recently. This adds a whole different level of complexity as the initial data for matter is now all of the allowed Hilbert Space. The recent analysis by Berczi et al has been done considering a quantum scalar field in a coherent state \cite{Berczi_JHEP,Berczi_PRD}. 

We have studied the dynamics of a test quantum field in the background spacetime undergoing gravitational collapse, adapting the numerical setup introduced for the critical collapse of a quantum field \cite{Berczi_JHEP,Berczi_PRD}. The mode functions for the test scalar field are initialized using the Minkowski modes at the start of the simulation, and the equations of motion for modes are evolved numerically along with the Einstein-Klein-Gordon system that gives an evolving background geometry. The vacuum of the quantum field is introduced at the start of the simulation using Minkowski modes, which gives the notion of the {\it in-state} (which is equivalent to the Unruh vacuum in standard parlance), and we investigate the dynamics of the {\it in-in correlations} of the test quantum field where the background geometry is undergoing critical collapse. We consider two cases of background evolution, supercritical and subcritical collapse, where the evolution leads to spacetime with and without black hole, respectively. We look at the dynamics of the equal time vacuum correlations of the scalar field and its conjugate momentum in those scenarios. For a near-critical scenario, the horizon appears after a long time and develops numerical instabilities, and we will consider this case in future analysis using a recently appeared {\it Julia}-based code \cite{Berczi:2024yhb}.

We find that the vacuum correlations of the field operator and its conjugate momentum develop non-local features across the horizon that represent the entangled Hawking partners, following the terminology in \cite{Schutzhold:2010ig}. The equal time vacuum correlations for the case where one point is inside the horizon $r'<r_{AH}$ demonstrate this behavior. Initially, there is a local peak at $r=r'$, that gets correlated to the peak outside the horizon as the black hole forms during the supercritical collapse, with the amplitude of the non-local peak being of the same order as that of the local peak. The location of the peak outside the horizon travels outwards as the system evolves during the stationary horizon regime, representing outgoing Hawking flux towards the asymptotic observer. This behavior is seen in vacuum correlations for cases of both the field operator and its conjugate momentum, but with varying oscillatory features. For scalar field correlations, the strong non-local features are masked by the apparent horizon, while the correlations for the momentum operator do have off-diagonal behavior outside the horizon that appears perpendicular to the main diagonal. 

For the quantum fields in Schwarzschild spacetime, the equal time correlations did not exhibit any non-local features across the horizon \cite{Balbinot:2021bnp}. This is attributed to the central singularity that absorbs the Hawking partner inside the horizon and sweeps away any characteristic peak in the exterior. The dynamics of the correlations for the set up under consideration closely resemble the toy model proposed in \cite{Schutzhold:2010ig}. The local dynamics for $r=r'$ inside the horizon is frozen due to the collapse of the lapse, leading to the stationary behavior of the diagonal correlations inside the horizon. Interestingly, the vanishing lapse does not prevent the non-local evolution, leading to a build-up of correlations across the horizon that indicates entangled pair of Hawking quanta, where the partner outside the horizon travels away from the horizon as the system evolves. In this picture, the quantum atmosphere is attributed to the location where the peak in correlations first appears \cite{Balbinot:2021bnp,Fontana:2023zqz}. The behavior of correlations indicates the existence of the quantum atmosphere as the correlation peak emerge well outside the location of apparent horizon.  It is interesting to see how the quantum field and its correlations will behave under different gauge choices for the lapse function.

The density correlations for the case of analogue black holes have some universal features: negative correlations along the diagonal $x=x'$ that indicate anti-bunching due to repulsive interaction, non-local correlations in the off-diagonal u-d sector where one point is in the subsonic (upstream) region and the other in the supersonic (downstream) region, and a pair of tongues originating from the horizon and in the supersonic sector which is the interior region of the analogue black hole \cite{Carusotto:2008ep}. The non-local correlations in the off-diagonal sector appear along the line perpendicular to the main diagonal and pass through the horizon $x=0=x'$, referred to as the {\it Hawking moustache} \cite{Berti:2024cut}, are termed the signature of Hawking radiation \cite{Carusotto:2008ep}. In the set up under consideration, the field correlations do develop non-local character in the in-ext region (where one point is in the interior region and the other is in the exterior) as the horizon appears, but they are of a different character as compared to the analogue black holes. This is expected as the two systems do not have exact equivalence, the interior and exterior regions in the analogue black holes are of infinite extent $x>0$ interior and $x<0$ exterior, whereas in our case, the interior maps to $r\in(0,r_{AH})$ and exterior maps to $r\in(r_{AH},\infty)$. Furthermore, instead of correlations of the scalar field and its conjugate momentum, correlations of the energy density operator of the quantum field would facilitate a more appropriate comparison. 

Coming to the question of information paradox, the central enigma is a pure initial state is observed to evolve to a mixed state in a supposedly \emph{unitary} process \cite{Hawking_1976}. An interesting insight to the issue is \cite{Page:1993wv,Kraus:1994by}, the \emph{subtle} correlations between the emitted Hawking quanta are neglected that might restore unitarity and could be enough to recover information about the initial state. In essence, the argument is the contributions of neglected off-diagonal terms in the `thermal' reduced density matrix can render it pure, although with the skepticism that small corrections to the leading order Hawking terms are not enough to recover unitarity. In a toy model of gravitational collapse \cite{Saini_2015}, the non-local correlations between the emitted particles across the horizon that corresponds to the off-diagonal terms of reduced density matrix are shown to grow to the order of leading order terms thereby showing the unitarity of the Hawking process. Our results are dynamical realization of this setup and we indeed observe the \emph{not so subtle} non-local correlations that can in principle help restore the information of initial pure state for the asymptotic observer. However, the next step is to investigate the reduced system obtained by tracing over the interior degrees of freedom and check if the entanglement entropy indeed follows the page curve \cite{Page:1993wv}. 

\section*{acknowledgements}

The research of SS and HSS is supported by the Core Research Grant CRG/2021/003053 from the Science and Engineering Research Board, India. We are thankful to the Inter-University Centre for Astronomy and Astrophysics (IUCAA) for providing the HPC facility on pegasus cluster under the project hpc2501018. We would like to thank Benjamin Berczi and Thanasis Giannakopoulos for the helpful comments on their code.


\appendix

\bibliography{main}

\begin{thebibliography}{105}%
\makeatletter
\providecommand \@ifxundefined [1]{%
 \@ifx{#1\undefined}
}%
\providecommand \@ifnum [1]{%
 \ifnum #1\expandafter \@firstoftwo
 \else \expandafter \@secondoftwo
 \fi
}%
\providecommand \@ifx [1]{%
 \ifx #1\expandafter \@firstoftwo
 \else \expandafter \@secondoftwo
 \fi
}%
\providecommand \natexlab [1]{#1}%
\providecommand \enquote  [1]{``#1''}%
\providecommand \bibnamefont  [1]{#1}%
\providecommand \bibfnamefont [1]{#1}%
\providecommand \citenamefont [1]{#1}%
\providecommand \href@noop [0]{\@secondoftwo}%
\providecommand \href [0]{\begingroup \@sanitize@url \@href}%
\providecommand \@href[1]{\@@startlink{#1}\@@href}%
\providecommand \@@href[1]{\endgroup#1\@@endlink}%
\providecommand \@sanitize@url [0]{\catcode `\\12\catcode `\$12\catcode `\&12\catcode `\#12\catcode `\^12\catcode `\_12\catcode `\%12\relax}%
\providecommand \@@startlink[1]{}%
\providecommand \@@endlink[0]{}%
\providecommand \url  [0]{\begingroup\@sanitize@url \@url }%
\providecommand \@url [1]{\endgroup\@href {#1}{\urlprefix }}%
\providecommand \urlprefix  [0]{URL }%
\providecommand \Eprint [0]{\href }%
\providecommand \doibase [0]{http://dx.doi.org/}%
\providecommand \selectlanguage [0]{\@gobble}%
\providecommand \bibinfo  [0]{\@secondoftwo}%
\providecommand \bibfield  [0]{\@secondoftwo}%
\providecommand \translation [1]{[#1]}%
\providecommand \BibitemOpen [0]{}%
\providecommand \bibitemStop [0]{}%
\providecommand \bibitemNoStop [0]{.\EOS\space}%
\providecommand \EOS [0]{\spacefactor3000\relax}%
\providecommand \BibitemShut  [1]{\csname bibitem#1\endcsname}%
\let\auto@bib@innerbib\@empty
\bibitem [{\citenamefont {Hawking}(1975)}]{Hawking1975}%
  \BibitemOpen
  \bibfield  {author} {\bibinfo {author} {\bibfnamefont {S.~W.}\ \bibnamefont {Hawking}},\ }\href {\doibase 10.1007/bf02345020} {\bibfield  {journal} {\bibinfo  {journal} {Communications In Mathematical Physics}\ }\textbf {\bibinfo {volume} {43}},\ \bibinfo {pages} {199} (\bibinfo {year} {1975})}\BibitemShut {NoStop}%
\bibitem [{\citenamefont {Wald}(1975)}]{Wald:1975kc}%
  \BibitemOpen
  \bibfield  {author} {\bibinfo {author} {\bibfnamefont {R.~M.}\ \bibnamefont {Wald}},\ }\href {\doibase 10.1007/BF01609863} {\bibfield  {journal} {\bibinfo  {journal} {Commun. Math. Phys.}\ }\textbf {\bibinfo {volume} {45}},\ \bibinfo {pages} {9} (\bibinfo {year} {1975})}\BibitemShut {NoStop}%
\bibitem [{\citenamefont {Unruh}(1976)}]{Unruh:1976db}%
  \BibitemOpen
  \bibfield  {author} {\bibinfo {author} {\bibfnamefont {W.~G.}\ \bibnamefont {Unruh}},\ }\href {\doibase 10.1103/PhysRevD.14.870} {\bibfield  {journal} {\bibinfo  {journal} {Phys. Rev. D}\ }\textbf {\bibinfo {volume} {14}},\ \bibinfo {pages} {870} (\bibinfo {year} {1976})}\BibitemShut {NoStop}%
\bibitem [{\citenamefont {Birrell}\ and\ \citenamefont {Davies}(1982)}]{Birrell:1982ix}%
  \BibitemOpen
  \bibfield  {author} {\bibinfo {author} {\bibfnamefont {N.~D.}\ \bibnamefont {Birrell}}\ and\ \bibinfo {author} {\bibfnamefont {P.~C.~W.}\ \bibnamefont {Davies}},\ }\href {\doibase 10.1017/CBO9780511622632} {\emph {\bibinfo {title} {{Quantum Fields in Curved Space}}}},\ Cambridge Monographs on Mathematical Physics\ (\bibinfo  {publisher} {Cambridge University Press},\ \bibinfo {address} {Cambridge, UK},\ \bibinfo {year} {1982})\BibitemShut {NoStop}%
\bibitem [{\citenamefont {Kay}\ and\ \citenamefont {Wald}(1991)}]{Kay:1988mu}%
  \BibitemOpen
  \bibfield  {author} {\bibinfo {author} {\bibfnamefont {B.~S.}\ \bibnamefont {Kay}}\ and\ \bibinfo {author} {\bibfnamefont {R.~M.}\ \bibnamefont {Wald}},\ }\href {\doibase 10.1016/0370-1573(91)90015-E} {\bibfield  {journal} {\bibinfo  {journal} {Phys. Rept.}\ }\textbf {\bibinfo {volume} {207}},\ \bibinfo {pages} {49} (\bibinfo {year} {1991})}\BibitemShut {NoStop}%
\bibitem [{\citenamefont {Fulling}\ and\ \citenamefont {Ruijsenaars}(1987)}]{FULLING1987135}%
  \BibitemOpen
  \bibfield  {author} {\bibinfo {author} {\bibfnamefont {S.}~\bibnamefont {Fulling}}\ and\ \bibinfo {author} {\bibfnamefont {S.}~\bibnamefont {Ruijsenaars}},\ }\href {\doibase https://doi.org/10.1016/0370-1573(87)90136-0} {\bibfield  {journal} {\bibinfo  {journal} {Physics Reports}\ }\textbf {\bibinfo {volume} {152}},\ \bibinfo {pages} {135} (\bibinfo {year} {1987})}\BibitemShut {NoStop}%
\bibitem [{\citenamefont {Wald}(1999)}]{Wald:1999xu}%
  \BibitemOpen
  \bibfield  {author} {\bibinfo {author} {\bibfnamefont {R.~M.}\ \bibnamefont {Wald}},\ }\href {\doibase 10.1088/0264-9381/16/12A/309} {\bibfield  {journal} {\bibinfo  {journal} {Class. Quant. Grav.}\ }\textbf {\bibinfo {volume} {16}},\ \bibinfo {pages} {A177} (\bibinfo {year} {1999})},\ \Eprint {http://arxiv.org/abs/gr-qc/9901033} {arXiv:gr-qc/9901033} \BibitemShut {NoStop}%
\bibitem [{\citenamefont {Visser}(2003)}]{Visser:2001kq}%
  \BibitemOpen
  \bibfield  {author} {\bibinfo {author} {\bibfnamefont {M.}~\bibnamefont {Visser}},\ }\href {\doibase 10.1142/S0218271803003190} {\bibfield  {journal} {\bibinfo  {journal} {Int. J. Mod. Phys. D}\ }\textbf {\bibinfo {volume} {12}},\ \bibinfo {pages} {649} (\bibinfo {year} {2003})},\ \Eprint {http://arxiv.org/abs/hep-th/0106111} {arXiv:hep-th/0106111} \BibitemShut {NoStop}%
\bibitem [{\citenamefont {Helfer}(2003)}]{Helfer:2003va}%
  \BibitemOpen
  \bibfield  {author} {\bibinfo {author} {\bibfnamefont {A.~D.}\ \bibnamefont {Helfer}},\ }\href {\doibase 10.1088/0034-4885/66/6/202} {\bibfield  {journal} {\bibinfo  {journal} {Rept. Prog. Phys.}\ }\textbf {\bibinfo {volume} {66}},\ \bibinfo {pages} {943} (\bibinfo {year} {2003})},\ \Eprint {http://arxiv.org/abs/gr-qc/0304042} {arXiv:gr-qc/0304042} \BibitemShut {NoStop}%
\bibitem [{\citenamefont {Padmanabhan}(2005)}]{Padmanabhan:2003gd}%
  \BibitemOpen
  \bibfield  {author} {\bibinfo {author} {\bibfnamefont {T.}~\bibnamefont {Padmanabhan}},\ }\href {\doibase 10.1016/j.physrep.2004.10.003} {\bibfield  {journal} {\bibinfo  {journal} {Phys. Rept.}\ }\textbf {\bibinfo {volume} {406}},\ \bibinfo {pages} {49} (\bibinfo {year} {2005})},\ \Eprint {http://arxiv.org/abs/gr-qc/0311036} {arXiv:gr-qc/0311036} \BibitemShut {NoStop}%
\bibitem [{\citenamefont {Fabbri}\ and\ \citenamefont {Navarro-Salas}(2005)}]{Fabbri:2005mw}%
  \BibitemOpen
  \bibfield  {author} {\bibinfo {author} {\bibfnamefont {A.}~\bibnamefont {Fabbri}}\ and\ \bibinfo {author} {\bibfnamefont {J.}~\bibnamefont {Navarro-Salas}},\ }\href {\doibase 10.1142/p378} {\emph {\bibinfo {title} {{Modeling black hole evaporation}}}}\ (\bibinfo  {publisher} {World Scientific},\ \bibinfo {address} {Singapore},\ \bibinfo {year} {2005})\BibitemShut {NoStop}%
\bibitem [{\citenamefont {Padmanabhan}(2010)}]{Padmanabhan:2009vy}%
  \BibitemOpen
  \bibfield  {author} {\bibinfo {author} {\bibfnamefont {T.}~\bibnamefont {Padmanabhan}},\ }\href {\doibase 10.1088/0034-4885/73/4/046901} {\bibfield  {journal} {\bibinfo  {journal} {Rept. Prog. Phys.}\ }\textbf {\bibinfo {volume} {73}},\ \bibinfo {pages} {046901} (\bibinfo {year} {2010})},\ \Eprint {http://arxiv.org/abs/0911.5004} {arXiv:0911.5004 [gr-qc]} \BibitemShut {NoStop}%
\bibitem [{\citenamefont {Visser}(2015)}]{Visser:2014ypa}%
  \BibitemOpen
  \bibfield  {author} {\bibinfo {author} {\bibfnamefont {M.}~\bibnamefont {Visser}},\ }\href {\doibase 10.1007/JHEP07(2015)009} {\bibfield  {journal} {\bibinfo  {journal} {JHEP}\ }\textbf {\bibinfo {volume} {07}},\ \bibinfo {pages} {009} (\bibinfo {year} {2015})},\ \Eprint {http://arxiv.org/abs/1409.7754} {arXiv:1409.7754 [gr-qc]} \BibitemShut {NoStop}%
\bibitem [{\citenamefont {Mathur}(2009)}]{Mathur:2009hf}%
  \BibitemOpen
  \bibfield  {author} {\bibinfo {author} {\bibfnamefont {S.~D.}\ \bibnamefont {Mathur}},\ }\href {\doibase 10.1088/0264-9381/26/22/224001} {\bibfield  {journal} {\bibinfo  {journal} {Class. Quant. Grav.}\ }\textbf {\bibinfo {volume} {26}},\ \bibinfo {pages} {224001} (\bibinfo {year} {2009})},\ \Eprint {http://arxiv.org/abs/0909.1038} {arXiv:0909.1038 [hep-th]} \BibitemShut {NoStop}%
\bibitem [{\citenamefont {Raju}(2022)}]{Raju:2020smc}%
  \BibitemOpen
  \bibfield  {author} {\bibinfo {author} {\bibfnamefont {S.}~\bibnamefont {Raju}},\ }\href {\doibase 10.1016/j.physrep.2021.10.001} {\bibfield  {journal} {\bibinfo  {journal} {Phys. Rept.}\ }\textbf {\bibinfo {volume} {943}},\ \bibinfo {pages} {1} (\bibinfo {year} {2022})},\ \Eprint {http://arxiv.org/abs/2012.05770} {arXiv:2012.05770 [hep-th]} \BibitemShut {NoStop}%
\bibitem [{\citenamefont {Harlow}(2016)}]{Harlow_2016}%
  \BibitemOpen
  \bibfield  {author} {\bibinfo {author} {\bibfnamefont {D.}~\bibnamefont {Harlow}},\ }\href {\doibase 10.1103/revmodphys.88.015002} {\bibfield  {journal} {\bibinfo  {journal} {Reviews of Modern Physics}\ }\textbf {\bibinfo {volume} {88}} (\bibinfo {year} {2016}),\ 10.1103/revmodphys.88.015002}\BibitemShut {NoStop}%
\bibitem [{\citenamefont {Unruh}\ and\ \citenamefont {Wald}(2017)}]{Unruh:2017uaw}%
  \BibitemOpen
  \bibfield  {author} {\bibinfo {author} {\bibfnamefont {W.~G.}\ \bibnamefont {Unruh}}\ and\ \bibinfo {author} {\bibfnamefont {R.~M.}\ \bibnamefont {Wald}},\ }\href {\doibase 10.1088/1361-6633/aa778e} {\bibfield  {journal} {\bibinfo  {journal} {Rept. Prog. Phys.}\ }\textbf {\bibinfo {volume} {80}},\ \bibinfo {pages} {092002} (\bibinfo {year} {2017})},\ \Eprint {http://arxiv.org/abs/1703.02140} {arXiv:1703.02140 [hep-th]} \BibitemShut {NoStop}%
\bibitem [{\citenamefont {Unruh}(1981)}]{Unruh:1980cg}%
  \BibitemOpen
  \bibfield  {author} {\bibinfo {author} {\bibfnamefont {W.~G.}\ \bibnamefont {Unruh}},\ }\href {\doibase 10.1103/PhysRevLett.46.1351} {\bibfield  {journal} {\bibinfo  {journal} {Phys. Rev. Lett.}\ }\textbf {\bibinfo {volume} {46}},\ \bibinfo {pages} {1351} (\bibinfo {year} {1981})}\BibitemShut {NoStop}%
\bibitem [{\citenamefont {Hawking}(1976)}]{Hawking_1976}%
  \BibitemOpen
  \bibfield  {author} {\bibinfo {author} {\bibfnamefont {S.~W.}\ \bibnamefont {Hawking}},\ }\href {\doibase 10.1103/PhysRevD.14.2460} {\bibfield  {journal} {\bibinfo  {journal} {Phys. Rev. D}\ }\textbf {\bibinfo {volume} {14}},\ \bibinfo {pages} {2460} (\bibinfo {year} {1976})}\BibitemShut {NoStop}%
\bibitem [{\citenamefont {Page}(1993{\natexlab{a}})}]{Page:1993df}%
  \BibitemOpen
  \bibfield  {author} {\bibinfo {author} {\bibfnamefont {D.~N.}\ \bibnamefont {Page}},\ }\href {\doibase 10.1103/PhysRevLett.71.1291} {\bibfield  {journal} {\bibinfo  {journal} {Phys. Rev. Lett.}\ }\textbf {\bibinfo {volume} {71}},\ \bibinfo {pages} {1291} (\bibinfo {year} {1993}{\natexlab{a}})},\ \Eprint {http://arxiv.org/abs/gr-qc/9305007} {arXiv:gr-qc/9305007} \BibitemShut {NoStop}%
\bibitem [{\citenamefont {Page}(1993{\natexlab{b}})}]{Page:1993wv}%
  \BibitemOpen
  \bibfield  {author} {\bibinfo {author} {\bibfnamefont {D.~N.}\ \bibnamefont {Page}},\ }\href {\doibase 10.1103/PhysRevLett.71.3743} {\bibfield  {journal} {\bibinfo  {journal} {Phys. Rev. Lett.}\ }\textbf {\bibinfo {volume} {71}},\ \bibinfo {pages} {3743} (\bibinfo {year} {1993}{\natexlab{b}})},\ \Eprint {http://arxiv.org/abs/hep-th/9306083} {arXiv:hep-th/9306083} \BibitemShut {NoStop}%
\bibitem [{\citenamefont {Kraus}\ and\ \citenamefont {Wilczek}(1995)}]{Kraus:1994by}%
  \BibitemOpen
  \bibfield  {author} {\bibinfo {author} {\bibfnamefont {P.}~\bibnamefont {Kraus}}\ and\ \bibinfo {author} {\bibfnamefont {F.}~\bibnamefont {Wilczek}},\ }\href {\doibase 10.1016/0550-3213(94)00411-7} {\bibfield  {journal} {\bibinfo  {journal} {Nucl. Phys. B}\ }\textbf {\bibinfo {volume} {433}},\ \bibinfo {pages} {403} (\bibinfo {year} {1995})},\ \Eprint {http://arxiv.org/abs/gr-qc/9408003} {arXiv:gr-qc/9408003} \BibitemShut {NoStop}%
\bibitem [{\citenamefont {Chakraborty}\ and\ \citenamefont {Lochan}(2017)}]{Chakraborty_2017}%
  \BibitemOpen
  \bibfield  {author} {\bibinfo {author} {\bibfnamefont {S.}~\bibnamefont {Chakraborty}}\ and\ \bibinfo {author} {\bibfnamefont {K.}~\bibnamefont {Lochan}},\ }\href {\doibase 10.3390/universe3030055} {\bibfield  {journal} {\bibinfo  {journal} {Universe}\ }\textbf {\bibinfo {volume} {3}},\ \bibinfo {pages} {55} (\bibinfo {year} {2017})}\BibitemShut {NoStop}%
\bibitem [{\citenamefont {Almheiri}\ \emph {et~al.}(2020)\citenamefont {Almheiri}, \citenamefont {Mahajan}, \citenamefont {Maldacena},\ and\ \citenamefont {Zhao}}]{Almheiri_2020}%
  \BibitemOpen
  \bibfield  {author} {\bibinfo {author} {\bibfnamefont {A.}~\bibnamefont {Almheiri}}, \bibinfo {author} {\bibfnamefont {R.}~\bibnamefont {Mahajan}}, \bibinfo {author} {\bibfnamefont {J.}~\bibnamefont {Maldacena}}, \ and\ \bibinfo {author} {\bibfnamefont {Y.}~\bibnamefont {Zhao}},\ }\href {\doibase 10.1007/jhep03(2020)149} {\bibfield  {journal} {\bibinfo  {journal} {Journal of High Energy Physics}\ }\textbf {\bibinfo {volume} {2020}} (\bibinfo {year} {2020}),\ 10.1007/jhep03(2020)149}\BibitemShut {NoStop}%
\bibitem [{\citenamefont {Almheiri}\ \emph {et~al.}(2021)\citenamefont {Almheiri}, \citenamefont {Hartman}, \citenamefont {Maldacena}, \citenamefont {Shaghoulian},\ and\ \citenamefont {Tajdini}}]{Almheiri:2020cfm}%
  \BibitemOpen
  \bibfield  {author} {\bibinfo {author} {\bibfnamefont {A.}~\bibnamefont {Almheiri}}, \bibinfo {author} {\bibfnamefont {T.}~\bibnamefont {Hartman}}, \bibinfo {author} {\bibfnamefont {J.}~\bibnamefont {Maldacena}}, \bibinfo {author} {\bibfnamefont {E.}~\bibnamefont {Shaghoulian}}, \ and\ \bibinfo {author} {\bibfnamefont {A.}~\bibnamefont {Tajdini}},\ }\href {\doibase 10.1103/RevModPhys.93.035002} {\bibfield  {journal} {\bibinfo  {journal} {Rev. Mod. Phys.}\ }\textbf {\bibinfo {volume} {93}},\ \bibinfo {pages} {035002} (\bibinfo {year} {2021})},\ \Eprint {http://arxiv.org/abs/2006.06872} {arXiv:2006.06872 [hep-th]} \BibitemShut {NoStop}%
\bibitem [{\citenamefont {Unruh}(1977)}]{Unruh_atmosphere}%
  \BibitemOpen
  \bibfield  {author} {\bibinfo {author} {\bibfnamefont {W.~G.}\ \bibnamefont {Unruh}},\ }\href {\doibase 10.1103/PhysRevD.15.365} {\bibfield  {journal} {\bibinfo  {journal} {Phys. Rev. D}\ }\textbf {\bibinfo {volume} {15}},\ \bibinfo {pages} {365} (\bibinfo {year} {1977})}\BibitemShut {NoStop}%
\bibitem [{\citenamefont {Schutzhold}\ and\ \citenamefont {Unruh}(2008)}]{Schutzhold:2008tx}%
  \BibitemOpen
  \bibfield  {author} {\bibinfo {author} {\bibfnamefont {R.}~\bibnamefont {Schutzhold}}\ and\ \bibinfo {author} {\bibfnamefont {W.~G.}\ \bibnamefont {Unruh}},\ }\href {\doibase 10.1103/PhysRevD.78.041504} {\bibfield  {journal} {\bibinfo  {journal} {Phys. Rev. D}\ }\textbf {\bibinfo {volume} {78}},\ \bibinfo {pages} {041504} (\bibinfo {year} {2008})},\ \Eprint {http://arxiv.org/abs/0804.1686} {arXiv:0804.1686 [gr-qc]} \BibitemShut {NoStop}%
\bibitem [{\citenamefont {Schutzhold}\ and\ \citenamefont {Unruh}(2010)}]{Schutzhold:2010ig}%
  \BibitemOpen
  \bibfield  {author} {\bibinfo {author} {\bibfnamefont {R.}~\bibnamefont {Schutzhold}}\ and\ \bibinfo {author} {\bibfnamefont {W.~G.}\ \bibnamefont {Unruh}},\ }\href {\doibase 10.1103/PhysRevD.81.124033} {\bibfield  {journal} {\bibinfo  {journal} {Phys. Rev. D}\ }\textbf {\bibinfo {volume} {81}},\ \bibinfo {pages} {124033} (\bibinfo {year} {2010})},\ \Eprint {http://arxiv.org/abs/1002.1844} {arXiv:1002.1844 [gr-qc]} \BibitemShut {NoStop}%
\bibitem [{\citenamefont {Balbinot}\ and\ \citenamefont {Fabbri}(2022)}]{Balbinot:2021bnp}%
  \BibitemOpen
  \bibfield  {author} {\bibinfo {author} {\bibfnamefont {R.}~\bibnamefont {Balbinot}}\ and\ \bibinfo {author} {\bibfnamefont {A.}~\bibnamefont {Fabbri}},\ }\href {\doibase 10.1103/PhysRevD.105.045010} {\bibfield  {journal} {\bibinfo  {journal} {Phys. Rev. D}\ }\textbf {\bibinfo {volume} {105}},\ \bibinfo {pages} {045010} (\bibinfo {year} {2022})},\ \Eprint {http://arxiv.org/abs/2107.00702} {arXiv:2107.00702 [gr-qc]} \BibitemShut {NoStop}%
\bibitem [{\citenamefont {Fontana}\ and\ \citenamefont {Rinaldi}(2023)}]{Fontana:2023zqz}%
  \BibitemOpen
  \bibfield  {author} {\bibinfo {author} {\bibfnamefont {M.}~\bibnamefont {Fontana}}\ and\ \bibinfo {author} {\bibfnamefont {M.}~\bibnamefont {Rinaldi}},\ }\href {\doibase 10.1103/PhysRevD.108.125003} {\bibfield  {journal} {\bibinfo  {journal} {Phys. Rev. D}\ }\textbf {\bibinfo {volume} {108}},\ \bibinfo {pages} {125003} (\bibinfo {year} {2023})},\ \Eprint {http://arxiv.org/abs/2302.08804} {arXiv:2302.08804 [gr-qc]} \BibitemShut {NoStop}%
\bibitem [{\citenamefont {Kaczmarek}\ and\ \citenamefont {Szcz\k{e}\'{s}niak}(2024)}]{Kaczmarek:2023kpn}%
  \BibitemOpen
  \bibfield  {author} {\bibinfo {author} {\bibfnamefont {A.~Z.}\ \bibnamefont {Kaczmarek}}\ and\ \bibinfo {author} {\bibfnamefont {D.}~\bibnamefont {Szcz\k{e}\'{s}niak}},\ }\href {\doibase 10.1016/j.physletb.2023.138364} {\bibfield  {journal} {\bibinfo  {journal} {Phys. Lett. B}\ }\textbf {\bibinfo {volume} {848}},\ \bibinfo {pages} {138364} (\bibinfo {year} {2024})},\ \Eprint {http://arxiv.org/abs/2306.09941} {arXiv:2306.09941 [gr-qc]} \BibitemShut {NoStop}%
\bibitem [{\citenamefont {Giddings}(2016)}]{Giddings:2015uzr}%
  \BibitemOpen
  \bibfield  {author} {\bibinfo {author} {\bibfnamefont {S.~B.}\ \bibnamefont {Giddings}},\ }\href {\doibase 10.1016/j.physletb.2015.12.076} {\bibfield  {journal} {\bibinfo  {journal} {Phys. Lett. B}\ }\textbf {\bibinfo {volume} {754}},\ \bibinfo {pages} {39} (\bibinfo {year} {2016})},\ \Eprint {http://arxiv.org/abs/1511.08221} {arXiv:1511.08221 [hep-th]} \BibitemShut {NoStop}%
\bibitem [{\citenamefont {Dey}\ \emph {et~al.}(2017)\citenamefont {Dey}, \citenamefont {Liberati},\ and\ \citenamefont {Pranzetti}}]{Dey:2017yez}%
  \BibitemOpen
  \bibfield  {author} {\bibinfo {author} {\bibfnamefont {R.}~\bibnamefont {Dey}}, \bibinfo {author} {\bibfnamefont {S.}~\bibnamefont {Liberati}}, \ and\ \bibinfo {author} {\bibfnamefont {D.}~\bibnamefont {Pranzetti}},\ }\href {\doibase 10.1016/j.physletb.2017.09.076} {\bibfield  {journal} {\bibinfo  {journal} {Phys. Lett. B}\ }\textbf {\bibinfo {volume} {774}},\ \bibinfo {pages} {308} (\bibinfo {year} {2017})},\ \Eprint {http://arxiv.org/abs/1701.06161} {arXiv:1701.06161 [gr-qc]} \BibitemShut {NoStop}%
\bibitem [{\citenamefont {Dey}\ \emph {et~al.}(2019)\citenamefont {Dey}, \citenamefont {Liberati}, \citenamefont {Mirzaiyan},\ and\ \citenamefont {Pranzetti}}]{Dey:2019ugf}%
  \BibitemOpen
  \bibfield  {author} {\bibinfo {author} {\bibfnamefont {R.}~\bibnamefont {Dey}}, \bibinfo {author} {\bibfnamefont {S.}~\bibnamefont {Liberati}}, \bibinfo {author} {\bibfnamefont {Z.}~\bibnamefont {Mirzaiyan}}, \ and\ \bibinfo {author} {\bibfnamefont {D.}~\bibnamefont {Pranzetti}},\ }\href {\doibase 10.1016/j.physletb.2019.134828} {\bibfield  {journal} {\bibinfo  {journal} {Phys. Lett. B}\ }\textbf {\bibinfo {volume} {797}},\ \bibinfo {pages} {134828} (\bibinfo {year} {2019})},\ \Eprint {http://arxiv.org/abs/1906.02958} {arXiv:1906.02958 [gr-qc]} \BibitemShut {NoStop}%
\bibitem [{\citenamefont {Barcelo}\ \emph {et~al.}(2005)\citenamefont {Barcelo}, \citenamefont {Liberati},\ and\ \citenamefont {Visser}}]{Barcelo:2005fc}%
  \BibitemOpen
  \bibfield  {author} {\bibinfo {author} {\bibfnamefont {C.}~\bibnamefont {Barcelo}}, \bibinfo {author} {\bibfnamefont {S.}~\bibnamefont {Liberati}}, \ and\ \bibinfo {author} {\bibfnamefont {M.}~\bibnamefont {Visser}},\ }\href {\doibase 10.12942/lrr-2005-12} {\bibfield  {journal} {\bibinfo  {journal} {Living Rev. Rel.}\ }\textbf {\bibinfo {volume} {8}},\ \bibinfo {pages} {12} (\bibinfo {year} {2005})},\ \Eprint {http://arxiv.org/abs/gr-qc/0505065} {arXiv:gr-qc/0505065} \BibitemShut {NoStop}%
\bibitem [{\citenamefont {Novello}\ \emph {et~al.}(2002)\citenamefont {Novello}, \citenamefont {Visser},\ and\ \citenamefont {Volovik}}]{Novello2002-ix}%
  \BibitemOpen
  \bibfield  {author} {\bibinfo {author} {\bibfnamefont {M.}~\bibnamefont {Novello}}, \bibinfo {author} {\bibfnamefont {M.}~\bibnamefont {Visser}}, \ and\ \bibinfo {author} {\bibfnamefont {G.}~\bibnamefont {Volovik}},\ }\href@noop {} {\emph {\bibinfo {title} {Artificial Black Holes}}},\ edited by\ \bibinfo {editor} {\bibfnamefont {M.}~\bibnamefont {Novello}}, \bibinfo {editor} {\bibfnamefont {M.}~\bibnamefont {Visser}}, \ and\ \bibinfo {editor} {\bibfnamefont {G.~E.}\ \bibnamefont {Volovik}}\ (\bibinfo  {publisher} {World Scientific Publishing},\ \bibinfo {address} {Singapore, Singapore},\ \bibinfo {year} {2002})\BibitemShut {NoStop}%
\bibitem [{\citenamefont {Visser}(1998)}]{Visser:1997ux}%
  \BibitemOpen
  \bibfield  {author} {\bibinfo {author} {\bibfnamefont {M.}~\bibnamefont {Visser}},\ }\href {\doibase 10.1088/0264-9381/15/6/024} {\bibfield  {journal} {\bibinfo  {journal} {Class. Quant. Grav.}\ }\textbf {\bibinfo {volume} {15}},\ \bibinfo {pages} {1767} (\bibinfo {year} {1998})},\ \Eprint {http://arxiv.org/abs/gr-qc/9712010} {arXiv:gr-qc/9712010} \BibitemShut {NoStop}%
\bibitem [{\citenamefont {Jacobson}\ and\ \citenamefont {Volovik}(1998)}]{Jacobson:1998ms}%
  \BibitemOpen
  \bibfield  {author} {\bibinfo {author} {\bibfnamefont {T.~A.}\ \bibnamefont {Jacobson}}\ and\ \bibinfo {author} {\bibfnamefont {G.~E.}\ \bibnamefont {Volovik}},\ }\href {\doibase 10.1103/PhysRevD.58.064021} {\bibfield  {journal} {\bibinfo  {journal} {Phys. Rev. D}\ }\textbf {\bibinfo {volume} {58}},\ \bibinfo {pages} {064021} (\bibinfo {year} {1998})},\ \Eprint {http://arxiv.org/abs/cond-mat/9801308} {arXiv:cond-mat/9801308} \BibitemShut {NoStop}%
\bibitem [{\citenamefont {Leonhardt}\ and\ \citenamefont {Piwnicki}(2000)}]{Leonhardt:2000fd}%
  \BibitemOpen
  \bibfield  {author} {\bibinfo {author} {\bibfnamefont {U.}~\bibnamefont {Leonhardt}}\ and\ \bibinfo {author} {\bibfnamefont {P.}~\bibnamefont {Piwnicki}},\ }\href {\doibase 10.1103/PhysRevLett.84.822} {\bibfield  {journal} {\bibinfo  {journal} {Phys. Rev. Lett.}\ }\textbf {\bibinfo {volume} {84}},\ \bibinfo {pages} {822} (\bibinfo {year} {2000})},\ \Eprint {http://arxiv.org/abs/cond-mat/9906332} {arXiv:cond-mat/9906332} \BibitemShut {NoStop}%
\bibitem [{\citenamefont {Unruh}\ and\ \citenamefont {Schutzhold}(2003)}]{Unruh:2003ss}%
  \BibitemOpen
  \bibfield  {author} {\bibinfo {author} {\bibfnamefont {W.~G.}\ \bibnamefont {Unruh}}\ and\ \bibinfo {author} {\bibfnamefont {R.}~\bibnamefont {Schutzhold}},\ }\href {\doibase 10.1103/PhysRevD.68.024008} {\bibfield  {journal} {\bibinfo  {journal} {Phys. Rev. D}\ }\textbf {\bibinfo {volume} {68}},\ \bibinfo {pages} {024008} (\bibinfo {year} {2003})},\ \Eprint {http://arxiv.org/abs/gr-qc/0303028} {arXiv:gr-qc/0303028} \BibitemShut {NoStop}%
\bibitem [{\citenamefont {Giovanazzi}\ \emph {et~al.}(2004)\citenamefont {Giovanazzi}, \citenamefont {Farrell}, \citenamefont {Kiss},\ and\ \citenamefont {Leonhardt}}]{Giovanazzi:2004pm}%
  \BibitemOpen
  \bibfield  {author} {\bibinfo {author} {\bibfnamefont {S.}~\bibnamefont {Giovanazzi}}, \bibinfo {author} {\bibfnamefont {C.}~\bibnamefont {Farrell}}, \bibinfo {author} {\bibfnamefont {T.}~\bibnamefont {Kiss}}, \ and\ \bibinfo {author} {\bibfnamefont {U.}~\bibnamefont {Leonhardt}},\ }\href {\doibase 10.1103/PhysRevA.70.063602} {\bibfield  {journal} {\bibinfo  {journal} {Phys. Rev. A}\ }\textbf {\bibinfo {volume} {70}},\ \bibinfo {pages} {063602} (\bibinfo {year} {2004})},\ \Eprint {http://arxiv.org/abs/cond-mat/0405007} {arXiv:cond-mat/0405007} \BibitemShut {NoStop}%
\bibitem [{\citenamefont {Schutzhold}\ and\ \citenamefont {Unruh}(2005)}]{Schutzhold:2004tv}%
  \BibitemOpen
  \bibfield  {author} {\bibinfo {author} {\bibfnamefont {R.}~\bibnamefont {Schutzhold}}\ and\ \bibinfo {author} {\bibfnamefont {W.~G.}\ \bibnamefont {Unruh}},\ }\href {\doibase 10.1103/PhysRevLett.95.031301} {\bibfield  {journal} {\bibinfo  {journal} {Phys. Rev. Lett.}\ }\textbf {\bibinfo {volume} {95}},\ \bibinfo {pages} {031301} (\bibinfo {year} {2005})},\ \Eprint {http://arxiv.org/abs/quant-ph/0408145} {arXiv:quant-ph/0408145} \BibitemShut {NoStop}%
\bibitem [{\citenamefont {Philbin}\ \emph {et~al.}(2008)\citenamefont {Philbin}, \citenamefont {Kuklewicz}, \citenamefont {Robertson}, \citenamefont {Hill}, \citenamefont {Konig},\ and\ \citenamefont {Leonhardt}}]{Philbin:2007ji}%
  \BibitemOpen
  \bibfield  {author} {\bibinfo {author} {\bibfnamefont {T.~G.}\ \bibnamefont {Philbin}}, \bibinfo {author} {\bibfnamefont {C.}~\bibnamefont {Kuklewicz}}, \bibinfo {author} {\bibfnamefont {S.}~\bibnamefont {Robertson}}, \bibinfo {author} {\bibfnamefont {S.}~\bibnamefont {Hill}}, \bibinfo {author} {\bibfnamefont {F.}~\bibnamefont {Konig}}, \ and\ \bibinfo {author} {\bibfnamefont {U.}~\bibnamefont {Leonhardt}},\ }\href {\doibase 10.1126/science.1153625} {\bibfield  {journal} {\bibinfo  {journal} {Science}\ }\textbf {\bibinfo {volume} {319}},\ \bibinfo {pages} {1367} (\bibinfo {year} {2008})},\ \Eprint {http://arxiv.org/abs/0711.4796} {arXiv:0711.4796 [gr-qc]} \BibitemShut {NoStop}%
\bibitem [{\citenamefont {Rousseaux}\ \emph {et~al.}(2008)\citenamefont {Rousseaux}, \citenamefont {Mathis}, \citenamefont {Maissa}, \citenamefont {Philbin},\ and\ \citenamefont {Leonhardt}}]{Rousseaux:2007is}%
  \BibitemOpen
  \bibfield  {author} {\bibinfo {author} {\bibfnamefont {G.}~\bibnamefont {Rousseaux}}, \bibinfo {author} {\bibfnamefont {C.}~\bibnamefont {Mathis}}, \bibinfo {author} {\bibfnamefont {P.}~\bibnamefont {Maissa}}, \bibinfo {author} {\bibfnamefont {T.~G.}\ \bibnamefont {Philbin}}, \ and\ \bibinfo {author} {\bibfnamefont {U.}~\bibnamefont {Leonhardt}},\ }\href {\doibase 10.1088/1367-2630/10/5/053015} {\bibfield  {journal} {\bibinfo  {journal} {New J. Phys.}\ }\textbf {\bibinfo {volume} {10}},\ \bibinfo {pages} {053015} (\bibinfo {year} {2008})},\ \Eprint {http://arxiv.org/abs/0711.4767} {arXiv:0711.4767 [gr-qc]} \BibitemShut {NoStop}%
\bibitem [{\citenamefont {Weinfurtner}\ \emph {et~al.}(2011)\citenamefont {Weinfurtner}, \citenamefont {Tedford}, \citenamefont {Penrice}, \citenamefont {Unruh},\ and\ \citenamefont {Lawrence}}]{Weinfurtner:2010nu}%
  \BibitemOpen
  \bibfield  {author} {\bibinfo {author} {\bibfnamefont {S.}~\bibnamefont {Weinfurtner}}, \bibinfo {author} {\bibfnamefont {E.~W.}\ \bibnamefont {Tedford}}, \bibinfo {author} {\bibfnamefont {M.~C.~J.}\ \bibnamefont {Penrice}}, \bibinfo {author} {\bibfnamefont {W.~G.}\ \bibnamefont {Unruh}}, \ and\ \bibinfo {author} {\bibfnamefont {G.~A.}\ \bibnamefont {Lawrence}},\ }\href {\doibase 10.1103/PhysRevLett.106.021302} {\bibfield  {journal} {\bibinfo  {journal} {Phys. Rev. Lett.}\ }\textbf {\bibinfo {volume} {106}},\ \bibinfo {pages} {021302} (\bibinfo {year} {2011})},\ \Eprint {http://arxiv.org/abs/1008.1911} {arXiv:1008.1911 [gr-qc]} \BibitemShut {NoStop}%
\bibitem [{\citenamefont {Belgiorno}\ \emph {et~al.}(2010)\citenamefont {Belgiorno}, \citenamefont {Cacciatori}, \citenamefont {Clerici}, \citenamefont {Gorini}, \citenamefont {Ortenzi}, \citenamefont {Rizzi}, \citenamefont {Rubino}, \citenamefont {Sala},\ and\ \citenamefont {Faccio}}]{Belgiorno:2010wn}%
  \BibitemOpen
  \bibfield  {author} {\bibinfo {author} {\bibfnamefont {F.}~\bibnamefont {Belgiorno}}, \bibinfo {author} {\bibfnamefont {S.~L.}\ \bibnamefont {Cacciatori}}, \bibinfo {author} {\bibfnamefont {M.}~\bibnamefont {Clerici}}, \bibinfo {author} {\bibfnamefont {V.}~\bibnamefont {Gorini}}, \bibinfo {author} {\bibfnamefont {G.}~\bibnamefont {Ortenzi}}, \bibinfo {author} {\bibfnamefont {L.}~\bibnamefont {Rizzi}}, \bibinfo {author} {\bibfnamefont {E.}~\bibnamefont {Rubino}}, \bibinfo {author} {\bibfnamefont {V.~G.}\ \bibnamefont {Sala}}, \ and\ \bibinfo {author} {\bibfnamefont {D.}~\bibnamefont {Faccio}},\ }\href {\doibase 10.1103/PhysRevLett.105.203901} {\bibfield  {journal} {\bibinfo  {journal} {Phys. Rev. Lett.}\ }\textbf {\bibinfo {volume} {105}},\ \bibinfo {pages} {203901} (\bibinfo {year} {2010})},\ \Eprint {http://arxiv.org/abs/1009.4634} {arXiv:1009.4634 [gr-qc]} \BibitemShut {NoStop}%
\bibitem [{\citenamefont {Horstmann}\ \emph {et~al.}(2010)\citenamefont {Horstmann}, \citenamefont {Reznik}, \citenamefont {Fagnocchi},\ and\ \citenamefont {Cirac}}]{Horstmann_2010}%
  \BibitemOpen
  \bibfield  {author} {\bibinfo {author} {\bibfnamefont {B.}~\bibnamefont {Horstmann}}, \bibinfo {author} {\bibfnamefont {B.}~\bibnamefont {Reznik}}, \bibinfo {author} {\bibfnamefont {S.}~\bibnamefont {Fagnocchi}}, \ and\ \bibinfo {author} {\bibfnamefont {J.~I.}\ \bibnamefont {Cirac}},\ }\href {\doibase 10.1103/PhysRevLett.104.250403} {\bibfield  {journal} {\bibinfo  {journal} {Phys. Rev. Lett.}\ }\textbf {\bibinfo {volume} {104}},\ \bibinfo {pages} {250403} (\bibinfo {year} {2010})}\BibitemShut {NoStop}%
\bibitem [{\citenamefont {Lombardo}\ and\ \citenamefont {Turiaci}(2012)}]{Lombardo_2012}%
  \BibitemOpen
  \bibfield  {author} {\bibinfo {author} {\bibfnamefont {F.~C.}\ \bibnamefont {Lombardo}}\ and\ \bibinfo {author} {\bibfnamefont {G.~J.}\ \bibnamefont {Turiaci}},\ }\href {\doibase 10.1103/physrevlett.108.261301} {\bibfield  {journal} {\bibinfo  {journal} {Physical Review Letters}\ }\textbf {\bibinfo {volume} {108}} (\bibinfo {year} {2012}),\ 10.1103/physrevlett.108.261301}\BibitemShut {NoStop}%
\bibitem [{\citenamefont {Busch}\ and\ \citenamefont {Parentani}(2014)}]{Busch_2014}%
  \BibitemOpen
  \bibfield  {author} {\bibinfo {author} {\bibfnamefont {X.}~\bibnamefont {Busch}}\ and\ \bibinfo {author} {\bibfnamefont {R.}~\bibnamefont {Parentani}},\ }\href {\doibase 10.1103/physrevd.89.105024} {\bibfield  {journal} {\bibinfo  {journal} {Physical Review D}\ }\textbf {\bibinfo {volume} {89}} (\bibinfo {year} {2014}),\ 10.1103/physrevd.89.105024}\BibitemShut {NoStop}%
\bibitem [{\citenamefont {Jacquet}\ \emph {et~al.}(2022)\citenamefont {Jacquet}, \citenamefont {Joly}, \citenamefont {Giacomelli}, \citenamefont {Claude}, \citenamefont {Glorieux}, \citenamefont {Bramati}, \citenamefont {Carusotto},\ and\ \citenamefont {Giacobino}}]{Jacquet:2022vak}%
  \BibitemOpen
  \bibfield  {author} {\bibinfo {author} {\bibfnamefont {M.~J.}\ \bibnamefont {Jacquet}}, \bibinfo {author} {\bibfnamefont {M.}~\bibnamefont {Joly}}, \bibinfo {author} {\bibfnamefont {L.}~\bibnamefont {Giacomelli}}, \bibinfo {author} {\bibfnamefont {F.}~\bibnamefont {Claude}}, \bibinfo {author} {\bibfnamefont {Q.}~\bibnamefont {Glorieux}}, \bibinfo {author} {\bibfnamefont {A.}~\bibnamefont {Bramati}}, \bibinfo {author} {\bibfnamefont {I.}~\bibnamefont {Carusotto}}, \ and\ \bibinfo {author} {\bibfnamefont {E.}~\bibnamefont {Giacobino}},\ }\href {\doibase 10.1140/epjd/s10053-022-00477-5} {\bibfield  {journal} {\bibinfo  {journal} {Eur. Phys. J. D}\ }\textbf {\bibinfo {volume} {76}},\ \bibinfo {pages} {152} (\bibinfo {year} {2022})},\ \bibinfo {note} {[Erratum: Eur.Phys.J.D 76, 247 (2022)]},\ \Eprint {http://arxiv.org/abs/2201.02038} {arXiv:2201.02038 [quant-ph]} \BibitemShut {NoStop}%
\bibitem [{\citenamefont {Jacquet}\ \emph {et~al.}(2023)\citenamefont {Jacquet}, \citenamefont {Giacomelli}, \citenamefont {Valnais}, \citenamefont {Joly}, \citenamefont {Claude}, \citenamefont {Giacobino}, \citenamefont {Glorieux}, \citenamefont {Carusotto},\ and\ \citenamefont {Bramati}}]{PhysRevLett.130.111501}%
  \BibitemOpen
  \bibfield  {author} {\bibinfo {author} {\bibfnamefont {M.~J.}\ \bibnamefont {Jacquet}}, \bibinfo {author} {\bibfnamefont {L.}~\bibnamefont {Giacomelli}}, \bibinfo {author} {\bibfnamefont {Q.}~\bibnamefont {Valnais}}, \bibinfo {author} {\bibfnamefont {M.}~\bibnamefont {Joly}}, \bibinfo {author} {\bibfnamefont {F.}~\bibnamefont {Claude}}, \bibinfo {author} {\bibfnamefont {E.}~\bibnamefont {Giacobino}}, \bibinfo {author} {\bibfnamefont {Q.}~\bibnamefont {Glorieux}}, \bibinfo {author} {\bibfnamefont {I.}~\bibnamefont {Carusotto}}, \ and\ \bibinfo {author} {\bibfnamefont {A.}~\bibnamefont {Bramati}},\ }\href {\doibase 10.1103/PhysRevLett.130.111501} {\bibfield  {journal} {\bibinfo  {journal} {Phys. Rev. Lett.}\ }\textbf {\bibinfo {volume} {130}},\ \bibinfo {pages} {111501} (\bibinfo {year} {2023})}\BibitemShut {NoStop}%
\bibitem [{\citenamefont {Berti}\ \emph {et~al.}(2025)\citenamefont {Berti}, \citenamefont {Fernandes}, \citenamefont {Butera}, \citenamefont {Recati}, \citenamefont {Wouters},\ and\ \citenamefont {Carusotto}}]{Berti:2024cut}%
  \BibitemOpen
  \bibfield  {author} {\bibinfo {author} {\bibfnamefont {A.}~\bibnamefont {Berti}}, \bibinfo {author} {\bibfnamefont {L.}~\bibnamefont {Fernandes}}, \bibinfo {author} {\bibfnamefont {S.~G.}\ \bibnamefont {Butera}}, \bibinfo {author} {\bibfnamefont {A.}~\bibnamefont {Recati}}, \bibinfo {author} {\bibfnamefont {M.}~\bibnamefont {Wouters}}, \ and\ \bibinfo {author} {\bibfnamefont {I.}~\bibnamefont {Carusotto}},\ }\href {\doibase 10.5802/crphys.226} {\bibfield  {journal} {\bibinfo  {journal} {Comptes Rendus Physique}\ }\textbf {\bibinfo {volume} {25}},\ \bibinfo {pages} {1} (\bibinfo {year} {2025})},\ \Eprint {http://arxiv.org/abs/2408.17292} {arXiv:2408.17292 [cond-mat.quant-gas]} \BibitemShut {NoStop}%
\bibitem [{\citenamefont {Garay}\ \emph {et~al.}(2000)\citenamefont {Garay}, \citenamefont {Anglin}, \citenamefont {Cirac},\ and\ \citenamefont {Zoller}}]{Garay:1999sk}%
  \BibitemOpen
  \bibfield  {author} {\bibinfo {author} {\bibfnamefont {L.~J.}\ \bibnamefont {Garay}}, \bibinfo {author} {\bibfnamefont {J.~R.}\ \bibnamefont {Anglin}}, \bibinfo {author} {\bibfnamefont {J.~I.}\ \bibnamefont {Cirac}}, \ and\ \bibinfo {author} {\bibfnamefont {P.}~\bibnamefont {Zoller}},\ }\href {\doibase 10.1103/PhysRevLett.85.4643} {\bibfield  {journal} {\bibinfo  {journal} {Phys. Rev. Lett.}\ }\textbf {\bibinfo {volume} {85}},\ \bibinfo {pages} {4643} (\bibinfo {year} {2000})},\ \Eprint {http://arxiv.org/abs/gr-qc/0002015} {arXiv:gr-qc/0002015} \BibitemShut {NoStop}%
\bibitem [{\citenamefont {Barcelo}\ \emph {et~al.}(2003)\citenamefont {Barcelo}, \citenamefont {Liberati},\ and\ \citenamefont {Visser}}]{Barcelo:2001ca}%
  \BibitemOpen
  \bibfield  {author} {\bibinfo {author} {\bibfnamefont {C.}~\bibnamefont {Barcelo}}, \bibinfo {author} {\bibfnamefont {S.}~\bibnamefont {Liberati}}, \ and\ \bibinfo {author} {\bibfnamefont {M.}~\bibnamefont {Visser}},\ }\href {\doibase 10.1142/S0217751X0301615X} {\bibfield  {journal} {\bibinfo  {journal} {Int. J. Mod. Phys. A}\ }\textbf {\bibinfo {volume} {18}},\ \bibinfo {pages} {3735} (\bibinfo {year} {2003})},\ \Eprint {http://arxiv.org/abs/gr-qc/0110036} {arXiv:gr-qc/0110036} \BibitemShut {NoStop}%
\bibitem [{\citenamefont {Giovanazzi}(2005)}]{Giovanazzi:2004zv}%
  \BibitemOpen
  \bibfield  {author} {\bibinfo {author} {\bibfnamefont {S.}~\bibnamefont {Giovanazzi}},\ }\href {\doibase 10.1103/PhysRevLett.94.061302} {\bibfield  {journal} {\bibinfo  {journal} {Phys. Rev. Lett.}\ }\textbf {\bibinfo {volume} {94}},\ \bibinfo {pages} {061302} (\bibinfo {year} {2005})},\ \Eprint {http://arxiv.org/abs/physics/0411064} {arXiv:physics/0411064} \BibitemShut {NoStop}%
\bibitem [{\citenamefont {Lahav}\ \emph {et~al.}(2010)\citenamefont {Lahav}, \citenamefont {Itah}, \citenamefont {Blumkin}, \citenamefont {Gordon},\ and\ \citenamefont {Steinhauer}}]{Lahav:2009wx}%
  \BibitemOpen
  \bibfield  {author} {\bibinfo {author} {\bibfnamefont {O.}~\bibnamefont {Lahav}}, \bibinfo {author} {\bibfnamefont {A.}~\bibnamefont {Itah}}, \bibinfo {author} {\bibfnamefont {A.}~\bibnamefont {Blumkin}}, \bibinfo {author} {\bibfnamefont {C.}~\bibnamefont {Gordon}}, \ and\ \bibinfo {author} {\bibfnamefont {J.}~\bibnamefont {Steinhauer}},\ }\href {\doibase 10.1103/PhysRevLett.105.240401} {\bibfield  {journal} {\bibinfo  {journal} {Phys. Rev. Lett.}\ }\textbf {\bibinfo {volume} {105}},\ \bibinfo {pages} {240401} (\bibinfo {year} {2010})},\ \Eprint {http://arxiv.org/abs/0906.1337} {arXiv:0906.1337 [cond-mat.quant-gas]} \BibitemShut {NoStop}%
\bibitem [{\citenamefont {Macher}\ and\ \citenamefont {Parentani}(2009)}]{Macher:2009nz}%
  \BibitemOpen
  \bibfield  {author} {\bibinfo {author} {\bibfnamefont {J.}~\bibnamefont {Macher}}\ and\ \bibinfo {author} {\bibfnamefont {R.}~\bibnamefont {Parentani}},\ }\href {\doibase 10.1103/PhysRevA.80.043601} {\bibfield  {journal} {\bibinfo  {journal} {Phys. Rev. A}\ }\textbf {\bibinfo {volume} {80}},\ \bibinfo {pages} {043601} (\bibinfo {year} {2009})},\ \Eprint {http://arxiv.org/abs/0905.3634} {arXiv:0905.3634 [cond-mat.quant-gas]} \BibitemShut {NoStop}%
\bibitem [{\citenamefont {Recati}\ \emph {et~al.}(2009)\citenamefont {Recati}, \citenamefont {Pavloff},\ and\ \citenamefont {Carusotto}}]{Recati:2009ya}%
  \BibitemOpen
  \bibfield  {author} {\bibinfo {author} {\bibfnamefont {A.}~\bibnamefont {Recati}}, \bibinfo {author} {\bibfnamefont {N.}~\bibnamefont {Pavloff}}, \ and\ \bibinfo {author} {\bibfnamefont {I.}~\bibnamefont {Carusotto}},\ }\href {\doibase 10.1103/PhysRevA.80.043603} {\bibfield  {journal} {\bibinfo  {journal} {Phys. Rev. A}\ }\textbf {\bibinfo {volume} {80}},\ \bibinfo {pages} {043603} (\bibinfo {year} {2009})},\ \Eprint {http://arxiv.org/abs/0907.4305} {arXiv:0907.4305 [cond-mat.quant-gas]} \BibitemShut {NoStop}%
\bibitem [{\citenamefont {Giovanazzi}(2011)}]{Giovanazzi_2011}%
  \BibitemOpen
  \bibfield  {author} {\bibinfo {author} {\bibfnamefont {S.}~\bibnamefont {Giovanazzi}},\ }\href {\doibase 10.1103/physrevlett.106.011302} {\bibfield  {journal} {\bibinfo  {journal} {Physical Review Letters}\ }\textbf {\bibinfo {volume} {106}} (\bibinfo {year} {2011}),\ 10.1103/physrevlett.106.011302}\BibitemShut {NoStop}%
\bibitem [{\citenamefont {Larr\'e}\ \emph {et~al.}(2012)\citenamefont {Larr\'e}, \citenamefont {Recati}, \citenamefont {Carusotto},\ and\ \citenamefont {Pavloff}}]{PhysRevA.85.013621}%
  \BibitemOpen
  \bibfield  {author} {\bibinfo {author} {\bibfnamefont {P.-E.}\ \bibnamefont {Larr\'e}}, \bibinfo {author} {\bibfnamefont {A.}~\bibnamefont {Recati}}, \bibinfo {author} {\bibfnamefont {I.}~\bibnamefont {Carusotto}}, \ and\ \bibinfo {author} {\bibfnamefont {N.}~\bibnamefont {Pavloff}},\ }\href {\doibase 10.1103/PhysRevA.85.013621} {\bibfield  {journal} {\bibinfo  {journal} {Phys. Rev. A}\ }\textbf {\bibinfo {volume} {85}},\ \bibinfo {pages} {013621} (\bibinfo {year} {2012})}\BibitemShut {NoStop}%
\bibitem [{\citenamefont {Finazzi}\ and\ \citenamefont {Carusotto}(2014)}]{Finazzi:2013sqa}%
  \BibitemOpen
  \bibfield  {author} {\bibinfo {author} {\bibfnamefont {S.}~\bibnamefont {Finazzi}}\ and\ \bibinfo {author} {\bibfnamefont {I.}~\bibnamefont {Carusotto}},\ }\href {\doibase 10.1103/PhysRevA.90.033607} {\bibfield  {journal} {\bibinfo  {journal} {Phys. Rev. A}\ }\textbf {\bibinfo {volume} {90}},\ \bibinfo {pages} {033607} (\bibinfo {year} {2014})},\ \Eprint {http://arxiv.org/abs/1309.3414} {arXiv:1309.3414 [cond-mat.quant-gas]} \BibitemShut {NoStop}%
\bibitem [{\citenamefont {Boiron}\ \emph {et~al.}(2015)\citenamefont {Boiron}, \citenamefont {Fabbri}, \citenamefont {Larr\'e}, \citenamefont {Pavloff}, \citenamefont {Westbrook},\ and\ \citenamefont {Zi\'n}}]{Boiron_2015}%
  \BibitemOpen
  \bibfield  {author} {\bibinfo {author} {\bibfnamefont {D.}~\bibnamefont {Boiron}}, \bibinfo {author} {\bibfnamefont {A.}~\bibnamefont {Fabbri}}, \bibinfo {author} {\bibfnamefont {P.~E.}\ \bibnamefont {Larr\'e}}, \bibinfo {author} {\bibfnamefont {N.}~\bibnamefont {Pavloff}}, \bibinfo {author} {\bibfnamefont {C.~I.}\ \bibnamefont {Westbrook}}, \ and\ \bibinfo {author} {\bibfnamefont {P.}~\bibnamefont {Zi\'n}},\ }\href {\doibase 10.1103/physrevlett.115.025301} {\bibfield  {journal} {\bibinfo  {journal} {Physical Review Letters}\ }\textbf {\bibinfo {volume} {115}} (\bibinfo {year} {2015}),\ 10.1103/physrevlett.115.025301}\BibitemShut {NoStop}%
\bibitem [{\citenamefont {Dudley}\ \emph {et~al.}(2018)\citenamefont {Dudley}, \citenamefont {Anderson}, \citenamefont {Balbinot},\ and\ \citenamefont {Fabbri}}]{Dudley:2018qpz}%
  \BibitemOpen
  \bibfield  {author} {\bibinfo {author} {\bibfnamefont {R.~A.}\ \bibnamefont {Dudley}}, \bibinfo {author} {\bibfnamefont {P.~R.}\ \bibnamefont {Anderson}}, \bibinfo {author} {\bibfnamefont {R.}~\bibnamefont {Balbinot}}, \ and\ \bibinfo {author} {\bibfnamefont {A.}~\bibnamefont {Fabbri}},\ }\href {\doibase 10.1103/PhysRevD.98.124011} {\bibfield  {journal} {\bibinfo  {journal} {Phys. Rev. D}\ }\textbf {\bibinfo {volume} {98}},\ \bibinfo {pages} {124011} (\bibinfo {year} {2018})},\ \Eprint {http://arxiv.org/abs/1809.03569} {arXiv:1809.03569 [gr-qc]} \BibitemShut {NoStop}%
\bibitem [{\citenamefont {Dudley}\ \emph {et~al.}(2020)\citenamefont {Dudley}, \citenamefont {Fabbri}, \citenamefont {Anderson},\ and\ \citenamefont {Balbinot}}]{Dudley:2020toe}%
  \BibitemOpen
  \bibfield  {author} {\bibinfo {author} {\bibfnamefont {R.~A.}\ \bibnamefont {Dudley}}, \bibinfo {author} {\bibfnamefont {A.}~\bibnamefont {Fabbri}}, \bibinfo {author} {\bibfnamefont {P.~R.}\ \bibnamefont {Anderson}}, \ and\ \bibinfo {author} {\bibfnamefont {R.}~\bibnamefont {Balbinot}},\ }\href {\doibase 10.1103/PhysRevD.102.105005} {\bibfield  {journal} {\bibinfo  {journal} {Phys. Rev. D}\ }\textbf {\bibinfo {volume} {102}},\ \bibinfo {pages} {105005} (\bibinfo {year} {2020})},\ \Eprint {http://arxiv.org/abs/2008.01433} {arXiv:2008.01433 [gr-qc]} \BibitemShut {NoStop}%
\bibitem [{\citenamefont {Palan}\ and\ \citenamefont {Wüster}(2022)}]{Palan_2022}%
  \BibitemOpen
  \bibfield  {author} {\bibinfo {author} {\bibfnamefont {Y.}~\bibnamefont {Palan}}\ and\ \bibinfo {author} {\bibfnamefont {S.}~\bibnamefont {Wüster}},\ }\href {\doibase 10.1103/physreva.106.053317} {\bibfield  {journal} {\bibinfo  {journal} {Physical Review A}\ }\textbf {\bibinfo {volume} {106}} (\bibinfo {year} {2022}),\ 10.1103/physreva.106.053317}\BibitemShut {NoStop}%
\bibitem [{\citenamefont {Anderson}\ \emph {et~al.}(2024)\citenamefont {Anderson}, \citenamefont {Balbinot}, \citenamefont {Dudley},\ and\ \citenamefont {Fabbri}}]{Anderson:2024fct}%
  \BibitemOpen
  \bibfield  {author} {\bibinfo {author} {\bibfnamefont {P.~R.}\ \bibnamefont {Anderson}}, \bibinfo {author} {\bibfnamefont {R.}~\bibnamefont {Balbinot}}, \bibinfo {author} {\bibfnamefont {R.~A.}\ \bibnamefont {Dudley}}, \ and\ \bibinfo {author} {\bibfnamefont {A.}~\bibnamefont {Fabbri}},\ }\href@noop {} {\enquote {\bibinfo {title} {{The peaks of the correlation function in acoustic black holes}},}\ } (\bibinfo {year} {2024}),\ \Eprint {http://arxiv.org/abs/2410.02700} {arXiv:2410.02700 [gr-qc]} \BibitemShut {NoStop}%
\bibitem [{\citenamefont {Hotta}\ \emph {et~al.}(2015)\citenamefont {Hotta}, \citenamefont {Sch\"utzhold},\ and\ \citenamefont {Unruh}}]{Hotta:2015yla}%
  \BibitemOpen
  \bibfield  {author} {\bibinfo {author} {\bibfnamefont {M.}~\bibnamefont {Hotta}}, \bibinfo {author} {\bibfnamefont {R.}~\bibnamefont {Sch\"utzhold}}, \ and\ \bibinfo {author} {\bibfnamefont {W.~G.}\ \bibnamefont {Unruh}},\ }\href {\doibase 10.1103/PhysRevD.91.124060} {\bibfield  {journal} {\bibinfo  {journal} {Phys. Rev. D}\ }\textbf {\bibinfo {volume} {91}},\ \bibinfo {pages} {124060} (\bibinfo {year} {2015})},\ \Eprint {http://arxiv.org/abs/1503.06109} {arXiv:1503.06109 [gr-qc]} \BibitemShut {NoStop}%
\bibitem [{\citenamefont {Balbinot}\ and\ \citenamefont {Fabbri}(2023)}]{Balbinot:2023grl}%
  \BibitemOpen
  \bibfield  {author} {\bibinfo {author} {\bibfnamefont {R.}~\bibnamefont {Balbinot}}\ and\ \bibinfo {author} {\bibfnamefont {A.}~\bibnamefont {Fabbri}},\ }\href {\doibase 10.1103/PhysRevD.108.045004} {\bibfield  {journal} {\bibinfo  {journal} {Phys. Rev. D}\ }\textbf {\bibinfo {volume} {108}},\ \bibinfo {pages} {045004} (\bibinfo {year} {2023})},\ \Eprint {http://arxiv.org/abs/2303.11039} {arXiv:2303.11039 [gr-qc]} \BibitemShut {NoStop}%
\bibitem [{\citenamefont {Balbinot}\ \emph {et~al.}(2008)\citenamefont {Balbinot}, \citenamefont {Fabbri}, \citenamefont {Fagnocchi}, \citenamefont {Recati},\ and\ \citenamefont {Carusotto}}]{Balbinot}%
  \BibitemOpen
  \bibfield  {author} {\bibinfo {author} {\bibfnamefont {R.}~\bibnamefont {Balbinot}}, \bibinfo {author} {\bibfnamefont {A.}~\bibnamefont {Fabbri}}, \bibinfo {author} {\bibfnamefont {S.}~\bibnamefont {Fagnocchi}}, \bibinfo {author} {\bibfnamefont {A.}~\bibnamefont {Recati}}, \ and\ \bibinfo {author} {\bibfnamefont {I.}~\bibnamefont {Carusotto}},\ }\href {\doibase 10.1103/PhysRevA.78.021603} {\bibfield  {journal} {\bibinfo  {journal} {Phys. Rev. A}\ }\textbf {\bibinfo {volume} {78}},\ \bibinfo {pages} {021603} (\bibinfo {year} {2008})}\BibitemShut {NoStop}%
\bibitem [{\citenamefont {Carusotto}\ \emph {et~al.}(2008)\citenamefont {Carusotto}, \citenamefont {Fagnocchi}, \citenamefont {Recati}, \citenamefont {Balbinot},\ and\ \citenamefont {Fabbri}}]{Carusotto:2008ep}%
  \BibitemOpen
  \bibfield  {author} {\bibinfo {author} {\bibfnamefont {I.}~\bibnamefont {Carusotto}}, \bibinfo {author} {\bibfnamefont {S.}~\bibnamefont {Fagnocchi}}, \bibinfo {author} {\bibfnamefont {A.}~\bibnamefont {Recati}}, \bibinfo {author} {\bibfnamefont {R.}~\bibnamefont {Balbinot}}, \ and\ \bibinfo {author} {\bibfnamefont {A.}~\bibnamefont {Fabbri}},\ }\href {\doibase 10.1088/1367-2630/10/10/103001} {\bibfield  {journal} {\bibinfo  {journal} {New J. Phys.}\ }\textbf {\bibinfo {volume} {10}},\ \bibinfo {pages} {103001} (\bibinfo {year} {2008})},\ \Eprint {http://arxiv.org/abs/0803.0507} {arXiv:0803.0507 [cond-mat.other]} \BibitemShut {NoStop}%
\bibitem [{\citenamefont {Michel}\ \emph {et~al.}(2016)\citenamefont {Michel}, \citenamefont {Coupechoux},\ and\ \citenamefont {Parentani}}]{Michel_2016}%
  \BibitemOpen
  \bibfield  {author} {\bibinfo {author} {\bibfnamefont {F.}~\bibnamefont {Michel}}, \bibinfo {author} {\bibfnamefont {J.-F. m.~c.}\ \bibnamefont {Coupechoux}}, \ and\ \bibinfo {author} {\bibfnamefont {R.}~\bibnamefont {Parentani}},\ }\href {\doibase 10.1103/PhysRevD.94.084027} {\bibfield  {journal} {\bibinfo  {journal} {Phys. Rev. D}\ }\textbf {\bibinfo {volume} {94}},\ \bibinfo {pages} {084027} (\bibinfo {year} {2016})}\BibitemShut {NoStop}%
\bibitem [{\citenamefont {Steinhauer}(2014)}]{Steinhauer:2014dra}%
  \BibitemOpen
  \bibfield  {author} {\bibinfo {author} {\bibfnamefont {J.}~\bibnamefont {Steinhauer}},\ }\href {\doibase 10.1038/NPHYS3104} {\bibfield  {journal} {\bibinfo  {journal} {Nature Phys.}\ }\textbf {\bibinfo {volume} {10}},\ \bibinfo {pages} {864} (\bibinfo {year} {2014})},\ \Eprint {http://arxiv.org/abs/1409.6550} {arXiv:1409.6550 [cond-mat.quant-gas]} \BibitemShut {NoStop}%
\bibitem [{\citenamefont {Steinhauer}(2015)}]{Steinhauer:2015ava}%
  \BibitemOpen
  \bibfield  {author} {\bibinfo {author} {\bibfnamefont {J.}~\bibnamefont {Steinhauer}},\ }\href {\doibase 10.1103/PhysRevD.92.024043} {\bibfield  {journal} {\bibinfo  {journal} {Phys. Rev. D}\ }\textbf {\bibinfo {volume} {92}},\ \bibinfo {pages} {024043} (\bibinfo {year} {2015})},\ \Eprint {http://arxiv.org/abs/1504.06583} {arXiv:1504.06583 [gr-qc]} \BibitemShut {NoStop}%
\bibitem [{\citenamefont {Steinhauer}(2016)}]{Steinhauer:2015saa}%
  \BibitemOpen
  \bibfield  {author} {\bibinfo {author} {\bibfnamefont {J.}~\bibnamefont {Steinhauer}},\ }\href {\doibase 10.1038/nphys3863} {\bibfield  {journal} {\bibinfo  {journal} {Nature Phys.}\ }\textbf {\bibinfo {volume} {12}},\ \bibinfo {pages} {959} (\bibinfo {year} {2016})},\ \Eprint {http://arxiv.org/abs/1510.00621} {arXiv:1510.00621 [gr-qc]} \BibitemShut {NoStop}%
\bibitem [{\citenamefont {Mu\~noz~de Nova}\ \emph {et~al.}(2019)\citenamefont {Mu\~noz~de Nova}, \citenamefont {Golubkov}, \citenamefont {Kolobov},\ and\ \citenamefont {Steinhauer}}]{MunozdeNova:2018fxv}%
  \BibitemOpen
  \bibfield  {author} {\bibinfo {author} {\bibfnamefont {J.~R.}\ \bibnamefont {Mu\~noz~de Nova}}, \bibinfo {author} {\bibfnamefont {K.}~\bibnamefont {Golubkov}}, \bibinfo {author} {\bibfnamefont {V.~I.}\ \bibnamefont {Kolobov}}, \ and\ \bibinfo {author} {\bibfnamefont {J.}~\bibnamefont {Steinhauer}},\ }\href {\doibase 10.1038/s41586-019-1241-0} {\bibfield  {journal} {\bibinfo  {journal} {Nature}\ }\textbf {\bibinfo {volume} {569}},\ \bibinfo {pages} {688} (\bibinfo {year} {2019})},\ \Eprint {http://arxiv.org/abs/1809.00913} {arXiv:1809.00913 [gr-qc]} \BibitemShut {NoStop}%
\bibitem [{\citenamefont {Kolobov}\ \emph {et~al.}(2021)\citenamefont {Kolobov}, \citenamefont {Golubkov}, \citenamefont {Mu\~noz~de Nova},\ and\ \citenamefont {Steinhauer}}]{Kolobov:2019qfs}%
  \BibitemOpen
  \bibfield  {author} {\bibinfo {author} {\bibfnamefont {V.~I.}\ \bibnamefont {Kolobov}}, \bibinfo {author} {\bibfnamefont {K.}~\bibnamefont {Golubkov}}, \bibinfo {author} {\bibfnamefont {J.~R.}\ \bibnamefont {Mu\~noz~de Nova}}, \ and\ \bibinfo {author} {\bibfnamefont {J.}~\bibnamefont {Steinhauer}},\ }\href {\doibase 10.1038/s41567-020-01076-0} {\bibfield  {journal} {\bibinfo  {journal} {Nature Phys.}\ }\textbf {\bibinfo {volume} {17}},\ \bibinfo {pages} {362} (\bibinfo {year} {2021})},\ \Eprint {http://arxiv.org/abs/1910.09363} {arXiv:1910.09363 [gr-qc]} \BibitemShut {NoStop}%
\bibitem [{\citenamefont {Fabbri}\ and\ \citenamefont {Balbinot}(2021)}]{Fabbri_2021}%
  \BibitemOpen
  \bibfield  {author} {\bibinfo {author} {\bibfnamefont {A.}~\bibnamefont {Fabbri}}\ and\ \bibinfo {author} {\bibfnamefont {R.}~\bibnamefont {Balbinot}},\ }\href {\doibase 10.1103/PhysRevLett.126.111301} {\bibfield  {journal} {\bibinfo  {journal} {Phys. Rev. Lett.}\ }\textbf {\bibinfo {volume} {126}},\ \bibinfo {pages} {111301} (\bibinfo {year} {2021})}\BibitemShut {NoStop}%
\bibitem [{\citenamefont {Braunstein}\ \emph {et~al.}(2013)\citenamefont {Braunstein}, \citenamefont {Pirandola},\ and\ \citenamefont {\.Zyczkowski}}]{Braunstein:2009my}%
  \BibitemOpen
  \bibfield  {author} {\bibinfo {author} {\bibfnamefont {S.~L.}\ \bibnamefont {Braunstein}}, \bibinfo {author} {\bibfnamefont {S.}~\bibnamefont {Pirandola}}, \ and\ \bibinfo {author} {\bibfnamefont {K.}~\bibnamefont {\.Zyczkowski}},\ }\href {\doibase 10.1103/PhysRevLett.110.101301} {\bibfield  {journal} {\bibinfo  {journal} {Phys. Rev. Lett.}\ }\textbf {\bibinfo {volume} {110}},\ \bibinfo {pages} {101301} (\bibinfo {year} {2013})},\ \Eprint {http://arxiv.org/abs/0907.1190} {arXiv:0907.1190 [quant-ph]} \BibitemShut {NoStop}%
\bibitem [{\citenamefont {Mathur}\ and\ \citenamefont {Plumberg}(2011)}]{Mathur_2011}%
  \BibitemOpen
  \bibfield  {author} {\bibinfo {author} {\bibfnamefont {S.~D.}\ \bibnamefont {Mathur}}\ and\ \bibinfo {author} {\bibfnamefont {C.~J.}\ \bibnamefont {Plumberg}},\ }\href {\doibase 10.1007/jhep09(2011)093} {\bibfield  {journal} {\bibinfo  {journal} {Journal of High Energy Physics}\ }\textbf {\bibinfo {volume} {2011}} (\bibinfo {year} {2011}),\ 10.1007/jhep09(2011)093}\BibitemShut {NoStop}%
\bibitem [{\citenamefont {Alonso-Serrano}\ and\ \citenamefont {Visser}(2018)}]{Alonso-Serrano:2015bcr}%
  \BibitemOpen
  \bibfield  {author} {\bibinfo {author} {\bibfnamefont {A.}~\bibnamefont {Alonso-Serrano}}\ and\ \bibinfo {author} {\bibfnamefont {M.}~\bibnamefont {Visser}},\ }\href {\doibase 10.1016/j.physletb.2017.11.020} {\bibfield  {journal} {\bibinfo  {journal} {Phys. Lett. B}\ }\textbf {\bibinfo {volume} {776}},\ \bibinfo {pages} {10} (\bibinfo {year} {2018})},\ \Eprint {http://arxiv.org/abs/1512.01890} {arXiv:1512.01890 [gr-qc]} \BibitemShut {NoStop}%
\bibitem [{\citenamefont {Alonso-Serrano}\ and\ \citenamefont {Visser}(2016)}]{Alonso_Serrano_2016}%
  \BibitemOpen
  \bibfield  {author} {\bibinfo {author} {\bibfnamefont {A.}~\bibnamefont {Alonso-Serrano}}\ and\ \bibinfo {author} {\bibfnamefont {M.}~\bibnamefont {Visser}},\ }\href {\doibase 10.1016/j.physletb.2016.04.023} {\bibfield  {journal} {\bibinfo  {journal} {Physics Letters B}\ }\textbf {\bibinfo {volume} {757}},\ \bibinfo {pages} {383–386} (\bibinfo {year} {2016})}\BibitemShut {NoStop}%
\bibitem [{\citenamefont {Lochan}\ and\ \citenamefont {Padmanabhan}(2016)}]{Lochan:PRL2015}%
  \BibitemOpen
  \bibfield  {author} {\bibinfo {author} {\bibfnamefont {K.}~\bibnamefont {Lochan}}\ and\ \bibinfo {author} {\bibfnamefont {T.}~\bibnamefont {Padmanabhan}},\ }\href {\doibase 10.1103/PhysRevLett.116.051301} {\bibfield  {journal} {\bibinfo  {journal} {Phys. Rev. Lett.}\ }\textbf {\bibinfo {volume} {116}},\ \bibinfo {pages} {051301} (\bibinfo {year} {2016})},\ \Eprint {http://arxiv.org/abs/1507.06402} {arXiv:1507.06402 [gr-qc]} \BibitemShut {NoStop}%
\bibitem [{\citenamefont {Lochan}\ \emph {et~al.}(2016)\citenamefont {Lochan}, \citenamefont {Chakraborty},\ and\ \citenamefont {Padmanabhan}}]{Lochan:PRD2016}%
  \BibitemOpen
  \bibfield  {author} {\bibinfo {author} {\bibfnamefont {K.}~\bibnamefont {Lochan}}, \bibinfo {author} {\bibfnamefont {S.}~\bibnamefont {Chakraborty}}, \ and\ \bibinfo {author} {\bibfnamefont {T.}~\bibnamefont {Padmanabhan}},\ }\href {\doibase 10.1103/PhysRevD.94.044056} {\bibfield  {journal} {\bibinfo  {journal} {Phys. Rev. D}\ }\textbf {\bibinfo {volume} {94}},\ \bibinfo {pages} {044056} (\bibinfo {year} {2016})},\ \Eprint {http://arxiv.org/abs/1604.04987} {arXiv:1604.04987 [gr-qc]} \BibitemShut {NoStop}%
\bibitem [{\citenamefont {Modak}\ \emph {et~al.}(2015)\citenamefont {Modak}, \citenamefont {Ortíz}, \citenamefont {Peña},\ and\ \citenamefont {Sudarsky}}]{Modak_2015}%
  \BibitemOpen
  \bibfield  {author} {\bibinfo {author} {\bibfnamefont {S.~K.}\ \bibnamefont {Modak}}, \bibinfo {author} {\bibfnamefont {L.}~\bibnamefont {Ortíz}}, \bibinfo {author} {\bibfnamefont {I.}~\bibnamefont {Peña}}, \ and\ \bibinfo {author} {\bibfnamefont {D.}~\bibnamefont {Sudarsky}},\ }\href {\doibase 10.1103/physrevd.91.124009} {\bibfield  {journal} {\bibinfo  {journal} {Physical Review D}\ }\textbf {\bibinfo {volume} {91}} (\bibinfo {year} {2015}),\ 10.1103/physrevd.91.124009}\BibitemShut {NoStop}%
\bibitem [{\citenamefont {Saini}\ and\ \citenamefont {Stojkovic}(2015)}]{Saini_2015}%
  \BibitemOpen
  \bibfield  {author} {\bibinfo {author} {\bibfnamefont {A.}~\bibnamefont {Saini}}\ and\ \bibinfo {author} {\bibfnamefont {D.}~\bibnamefont {Stojkovic}},\ }\href {\doibase 10.1103/physrevlett.114.111301} {\bibfield  {journal} {\bibinfo  {journal} {Physical Review Letters}\ }\textbf {\bibinfo {volume} {114}} (\bibinfo {year} {2015}),\ 10.1103/physrevlett.114.111301}\BibitemShut {NoStop}%
\bibitem [{\citenamefont {Christodoulou}(1986{\natexlab{a}})}]{Christodoulou:1986du}%
  \BibitemOpen
  \bibfield  {author} {\bibinfo {author} {\bibfnamefont {D.}~\bibnamefont {Christodoulou}},\ }\href {\doibase 10.1007/BF01463398} {\bibfield  {journal} {\bibinfo  {journal} {Commun. Math. Phys.}\ }\textbf {\bibinfo {volume} {106}},\ \bibinfo {pages} {587} (\bibinfo {year} {1986}{\natexlab{a}})}\BibitemShut {NoStop}%
\bibitem [{\citenamefont {Christodoulou}(1986{\natexlab{b}})}]{Christodoulou:1986zr}%
  \BibitemOpen
  \bibfield  {author} {\bibinfo {author} {\bibfnamefont {D.}~\bibnamefont {Christodoulou}},\ }\href {\doibase 10.1007/BF01205930} {\bibfield  {journal} {\bibinfo  {journal} {Commun. Math. Phys.}\ }\textbf {\bibinfo {volume} {105}},\ \bibinfo {pages} {337} (\bibinfo {year} {1986}{\natexlab{b}})}\BibitemShut {NoStop}%
\bibitem [{\citenamefont {Christodoulou}(1987{\natexlab{a}})}]{Christodoulou:1987vu}%
  \BibitemOpen
  \bibfield  {author} {\bibinfo {author} {\bibfnamefont {D.}~\bibnamefont {Christodoulou}},\ }\href {\doibase 10.1007/BF01208959} {\bibfield  {journal} {\bibinfo  {journal} {Commun. Math. Phys.}\ }\textbf {\bibinfo {volume} {109}},\ \bibinfo {pages} {591} (\bibinfo {year} {1987}{\natexlab{a}})}\BibitemShut {NoStop}%
\bibitem [{\citenamefont {Christodoulou}(1987{\natexlab{b}})}]{Christodoulou:1987vv}%
  \BibitemOpen
  \bibfield  {author} {\bibinfo {author} {\bibfnamefont {D.}~\bibnamefont {Christodoulou}},\ }\href {\doibase 10.1007/BF01208960} {\bibfield  {journal} {\bibinfo  {journal} {Commun. Math. Phys.}\ }\textbf {\bibinfo {volume} {109}},\ \bibinfo {pages} {613} (\bibinfo {year} {1987}{\natexlab{b}})}\BibitemShut {NoStop}%
\bibitem [{\citenamefont {Goldwirth}\ and\ \citenamefont {Piran}(1987)}]{Goldwirth:1987nu}%
  \BibitemOpen
  \bibfield  {author} {\bibinfo {author} {\bibfnamefont {D.~S.}\ \bibnamefont {Goldwirth}}\ and\ \bibinfo {author} {\bibfnamefont {T.}~\bibnamefont {Piran}},\ }\href {\doibase 10.1103/PhysRevD.36.3575} {\bibfield  {journal} {\bibinfo  {journal} {Phys. Rev. D}\ }\textbf {\bibinfo {volume} {36}},\ \bibinfo {pages} {3575} (\bibinfo {year} {1987})}\BibitemShut {NoStop}%
\bibitem [{\citenamefont {Choptuik}(1993)}]{Choptuik}%
  \BibitemOpen
  \bibfield  {author} {\bibinfo {author} {\bibfnamefont {M.~W.}\ \bibnamefont {Choptuik}},\ }\href {\doibase 10.1103/PhysRevLett.70.9} {\bibfield  {journal} {\bibinfo  {journal} {Phys. Rev. Lett.}\ }\textbf {\bibinfo {volume} {70}},\ \bibinfo {pages} {9} (\bibinfo {year} {1993})}\BibitemShut {NoStop}%
\bibitem [{\citenamefont {Lehner}(2001)}]{Lehner:2001wq}%
  \BibitemOpen
  \bibfield  {author} {\bibinfo {author} {\bibfnamefont {L.}~\bibnamefont {Lehner}},\ }\href {\doibase 10.1088/0264-9381/18/17/202} {\bibfield  {journal} {\bibinfo  {journal} {Class. Quant. Grav.}\ }\textbf {\bibinfo {volume} {18}},\ \bibinfo {pages} {R25} (\bibinfo {year} {2001})},\ \Eprint {http://arxiv.org/abs/gr-qc/0106072} {arXiv:gr-qc/0106072} \BibitemShut {NoStop}%
\bibitem [{\citenamefont {Gundlach}(2003)}]{Gundlach:2002sx}%
  \BibitemOpen
  \bibfield  {author} {\bibinfo {author} {\bibfnamefont {C.}~\bibnamefont {Gundlach}},\ }\href {\doibase 10.1016/S0370-1573(02)00560-4} {\bibfield  {journal} {\bibinfo  {journal} {Phys. Rept.}\ }\textbf {\bibinfo {volume} {376}},\ \bibinfo {pages} {339} (\bibinfo {year} {2003})},\ \Eprint {http://arxiv.org/abs/gr-qc/0210101} {arXiv:gr-qc/0210101} \BibitemShut {NoStop}%
\bibitem [{\citenamefont {Gundlach}\ and\ \citenamefont {Martin-Garcia}(2007)}]{Gundlach:2007gc}%
  \BibitemOpen
  \bibfield  {author} {\bibinfo {author} {\bibfnamefont {C.}~\bibnamefont {Gundlach}}\ and\ \bibinfo {author} {\bibfnamefont {J.~M.}\ \bibnamefont {Martin-Garcia}},\ }\href {\doibase 10.12942/lrr-2007-5} {\bibfield  {journal} {\bibinfo  {journal} {Living Rev. Rel.}\ }\textbf {\bibinfo {volume} {10}},\ \bibinfo {pages} {5} (\bibinfo {year} {2007})},\ \Eprint {http://arxiv.org/abs/0711.4620} {arXiv:0711.4620 [gr-qc]} \BibitemShut {NoStop}%
\bibitem [{\citenamefont {Alcubierre}(2012)}]{alcubierre}%
  \BibitemOpen
  \bibfield  {author} {\bibinfo {author} {\bibfnamefont {M.}~\bibnamefont {Alcubierre}},\ }\href@noop {} {\emph {\bibinfo {title} {Introduction to 3+1 Numerical Relativity}}}\ (\bibinfo  {publisher} {International Series of Monographs on Physics},\ \bibinfo {year} {2012})\BibitemShut {NoStop}%
\bibitem [{\citenamefont {Berczi}(2023)}]{BercziGH}%
  \BibitemOpen
  \bibfield  {author} {\bibinfo {author} {\bibfnamefont {B.}~\bibnamefont {Berczi}},\ }\href@noop {} {\enquote {\bibinfo {title} {{Gravitational collapse of quantum fields and Choptuik scaling}},}\ }\bibinfo {howpublished} {\url{https://github.com/benjibrcz/PhD}} (\bibinfo {year} {2023})\BibitemShut {NoStop}%
\bibitem [{\citenamefont {Berczi}\ \emph {et~al.}(2022)\citenamefont {Berczi}, \citenamefont {Saffin},\ and\ \citenamefont {Zhou}}]{Berczi_JHEP}%
  \BibitemOpen
  \bibfield  {author} {\bibinfo {author} {\bibfnamefont {B.}~\bibnamefont {Berczi}}, \bibinfo {author} {\bibfnamefont {P.~M.}\ \bibnamefont {Saffin}}, \ and\ \bibinfo {author} {\bibfnamefont {S.-Y.}\ \bibnamefont {Zhou}},\ }\href {\doibase 10.1007/JHEP02(2022)183} {\bibfield  {journal} {\bibinfo  {journal} {Journal of High Energy Physics}\ }\textbf {\bibinfo {volume} {2022}},\ \bibinfo {pages} {183} (\bibinfo {year} {2022})}\BibitemShut {NoStop}%
\bibitem [{\citenamefont {Berczi}\ \emph {et~al.}(2021)\citenamefont {Berczi}, \citenamefont {Saffin},\ and\ \citenamefont {Zhou}}]{Berczi_PRD}%
  \BibitemOpen
  \bibfield  {author} {\bibinfo {author} {\bibfnamefont {B.}~\bibnamefont {Berczi}}, \bibinfo {author} {\bibfnamefont {P.~M.}\ \bibnamefont {Saffin}}, \ and\ \bibinfo {author} {\bibfnamefont {S.-Y.}\ \bibnamefont {Zhou}},\ }\href {\doibase 10.1103/PhysRevD.104.L041703} {\bibfield  {journal} {\bibinfo  {journal} {Phys. Rev. D}\ }\textbf {\bibinfo {volume} {104}},\ \bibinfo {pages} {L041703} (\bibinfo {year} {2021})},\ \Eprint {http://arxiv.org/abs/2010.10142} {arXiv:2010.10142 [gr-qc]} \BibitemShut {NoStop}%
\bibitem [{\citenamefont {Guenther}\ \emph {et~al.}(2022)\citenamefont {Guenther}, \citenamefont {Hoelbling},\ and\ \citenamefont {Varnhorst}}]{Guenther}%
  \BibitemOpen
  \bibfield  {author} {\bibinfo {author} {\bibfnamefont {J.~N.}\ \bibnamefont {Guenther}}, \bibinfo {author} {\bibfnamefont {C.}~\bibnamefont {Hoelbling}}, \ and\ \bibinfo {author} {\bibfnamefont {L.}~\bibnamefont {Varnhorst}},\ }\href {\doibase 10.1103/PhysRevD.105.105010} {\bibfield  {journal} {\bibinfo  {journal} {Phys. Rev. D}\ }\textbf {\bibinfo {volume} {105}},\ \bibinfo {pages} {105010} (\bibinfo {year} {2022})},\ \Eprint {http://arxiv.org/abs/2010.13215} {arXiv:2010.13215 [gr-qc]} \BibitemShut {NoStop}%
\bibitem [{\citenamefont {Berczi}\ \emph {et~al.}(2024)\citenamefont {Berczi}, \citenamefont {Eriksson}, \citenamefont {Giannakopoulos},\ and\ \citenamefont {Saffin}}]{Berczi:2024yhb}%
  \BibitemOpen
  \bibfield  {author} {\bibinfo {author} {\bibfnamefont {B.}~\bibnamefont {Berczi}}, \bibinfo {author} {\bibfnamefont {M.}~\bibnamefont {Eriksson}}, \bibinfo {author} {\bibfnamefont {T.}~\bibnamefont {Giannakopoulos}}, \ and\ \bibinfo {author} {\bibfnamefont {P.~M.}\ \bibnamefont {Saffin}},\ }\href@noop {} {\enquote {\bibinfo {title} {{Quantum correlations in a gravitational collapse simulation with SpheriCo.jl}},}\ } (\bibinfo {year} {2024}),\ \Eprint {http://arxiv.org/abs/arXiv:2412.19722} {arXiv:arXiv:2412.19722 [gr-qc]} \BibitemShut {NoStop}%
\bibitem [{\citenamefont {Bona}\ \emph {et~al.}(1995)\citenamefont {Bona}, \citenamefont {Masso}, \citenamefont {Seidel},\ and\ \citenamefont {Stela}}]{BonaMasso}%
  \BibitemOpen
  \bibfield  {author} {\bibinfo {author} {\bibfnamefont {C.}~\bibnamefont {Bona}}, \bibinfo {author} {\bibfnamefont {J.}~\bibnamefont {Masso}}, \bibinfo {author} {\bibfnamefont {E.}~\bibnamefont {Seidel}}, \ and\ \bibinfo {author} {\bibfnamefont {J.}~\bibnamefont {Stela}},\ }\href@noop {} {\bibfield  {journal} {\bibinfo  {journal} {Phys. Rev. Lett.}\ }\textbf {\bibinfo {volume} {75}},\ \bibinfo {pages} {600} (\bibinfo {year} {1995})}\BibitemShut {NoStop}%
\bibitem [{\citenamefont {Pauli}\ and\ \citenamefont {Villars}(1949)}]{Pauli:1949zm}%
  \BibitemOpen
  \bibfield  {author} {\bibinfo {author} {\bibfnamefont {W.}~\bibnamefont {Pauli}}\ and\ \bibinfo {author} {\bibfnamefont {F.}~\bibnamefont {Villars}},\ }\href {\doibase 10.1103/RevModPhys.21.434} {\bibfield  {journal} {\bibinfo  {journal} {Rev. Mod. Phys.}\ ,\ \bibinfo {pages} {434}} (\bibinfo {year} {1949})}\BibitemShut {NoStop}%
\bibitem [{\citenamefont {Visser}(2018)}]{Visser:2016mtr}%
  \BibitemOpen
  \bibfield  {author} {\bibinfo {author} {\bibfnamefont {M.}~\bibnamefont {Visser}},\ }\href {\doibase 10.3390/particles1010010} {\bibfield  {journal} {\bibinfo  {journal} {Particles}\ }\textbf {\bibinfo {volume} {1}},\ \bibinfo {pages} {138} (\bibinfo {year} {2018})},\ \Eprint {http://arxiv.org/abs/1610.07264} {arXiv:1610.07264 [gr-qc]} \BibitemShut {NoStop}%
\bibitem [{\citenamefont {Kamenshchik}\ \emph {et~al.}(2018)\citenamefont {Kamenshchik}, \citenamefont {Starobinsky}, \citenamefont {Tronconi}, \citenamefont {Vardanyan},\ and\ \citenamefont {Venturi}}]{Kamenshchik:2018ttr}%
  \BibitemOpen
  \bibfield  {author} {\bibinfo {author} {\bibfnamefont {A.~Y.}\ \bibnamefont {Kamenshchik}}, \bibinfo {author} {\bibfnamefont {A.~A.}\ \bibnamefont {Starobinsky}}, \bibinfo {author} {\bibfnamefont {A.}~\bibnamefont {Tronconi}}, \bibinfo {author} {\bibfnamefont {T.}~\bibnamefont {Vardanyan}}, \ and\ \bibinfo {author} {\bibfnamefont {G.}~\bibnamefont {Venturi}},\ }\href {\doibase 10.1140/epjc/s10052-018-5703-6} {\bibfield  {journal} {\bibinfo  {journal} {Eur. Phys. J. C}\ }\textbf {\bibinfo {volume} {78}},\ \bibinfo {pages} {200} (\bibinfo {year} {2018})},\ \Eprint {http://arxiv.org/abs/1801.08434} {arXiv:1801.08434 [hep-th]} \BibitemShut {NoStop}%
\bibitem [{\citenamefont {Kreiss}\ and\ \citenamefont {Oliger}(1973)}]{kreiss1973methods}%
  \BibitemOpen
  \bibfield  {author} {\bibinfo {author} {\bibfnamefont {H.}~\bibnamefont {Kreiss}}\ and\ \bibinfo {author} {\bibfnamefont {J.}~\bibnamefont {Oliger}},\ }\href {https://books.google.co.in/books?id=wj9szwEACAAJ} {\emph {\bibinfo {title} {Methods for the approximate solution of time dependent problems}}},\ GARP Publications Series\ (\bibinfo  {publisher} {SELBSTVERL.) FEBR},\ \bibinfo {year} {1973})\BibitemShut {NoStop}%
\end{thebibliography}%

\end{document}